\documentclass[12pt]{article}

\usepackage{graphicx}

\input{epsf}

\def\rf#1{(\ref{eq:#1})}
\def\lab#1{\label{eq:#1}}

\def\br{\begin{eqnarray}}
\def\er{\end{eqnarray}}
\def\be{\begin{equation}}
\def\ee{\end{equation}}
\def\({\left(}
\def\){\right)}

\def\rlx{\relax\leavevmode}
\def\IR{\rlx\hbox{\rm I\kern-.18em R}}

\def\ve{\varepsilon}

\newcommand{\sbr}[2]{\left\lbrack\,{#1}\, ,\,{#2}\,\right\rbrack}

\def\IZ{\rlx\hbox{\sf Z\kern-.4em Z}}
\def\IR{\rlx\hbox{\rm I\kern-.18em R}}
\def\IC{\rlx\hbox{\,$\inbar\kern-.3em{\rm C}$}}
\def\one{\hbox{{1}\kern-.25em\hbox{l}}}

\begin{document}

\begin{titlepage}
\vspace*{-1cm}

\vskip 2cm

\vspace{.2in}
\begin{center}
{\large\bf Numerical and analytical tests of quasi-integrability in modified Sine-Gordon models}
\end{center}

\vspace{.5cm}

\begin{center}
L. A. Ferreira~$^{\star}$ and Wojtek J. Zakrzewski~$^{\dagger}$

\vspace{.3 in}
\small

\par \vskip .2in \noindent
$^{(\star)}$Instituto de F\'\i sica de S\~ao Carlos; IFSC/USP;\\
Universidade de S\~ao Paulo  \\ 
Caixa Postal 369, CEP 13560-970, S\~ao Carlos-SP, Brazil\\
email: laf@ifsc.usp.br

\par \vskip .2in \noindent
$^{(\dagger)}$~Department of Mathematical Sciences,\\
 University of Durham, Durham DH1 3LE, U.K.\\
email: W.J.Zakrzewski@durham.ac.uk

\normalsize
\end{center}


\begin{abstract}

Following our attempts to define quasi-integrability in which we related this concept to a particular symmetry of the two-soliton function we check this condition in three classes of modified Sine-Gordon models in (1+1) dimensions.
We find that the numerical results seen in various scatterings of two solitons and in the time evolution of breather-like structures support our ideas 
about the symmetry of the field configurations and its effects on the anomalies of the conservation laws of the charges.

\end{abstract} 
\end{titlepage}

\section{Introduction}
\label{sec:intro}
\setcounter{equation}{0}

In our recent work \cite{us} and \cite{us2} we have tried to define the concept 
of quasi-integrability for field theories in $(1+1)$ dimensions. The idea to introduce such a concept is based on 
observation that many processes described by field theories which are not integrable possess properties which are very close to those seen in integrable theories. Thus, for example, in many processes involving scattering of (soliton-like) extended structures these structures do not alter significantly their shapes and the whole scattering process generates very little radiation. In fact, after developing the concept of quasi-integrability in \cite{us,us2} we became aware of  previous observations of elastic scattering of solitons  in some non-integrable theories \cite{jarmo}, which reinforces the relevance of the development of such concepts.   

In our recent work we have looked in detail, both numerically
and analytically, at the scattering of solitons in two classes of models in (1+1) dimensions, namely the generalized models of the Sine Gordon type as introduced by Bazeia et al \cite{bazeia} and the generalized models of the Nonlinear Schr\"odinger type (in which the usual NLS potential $\vert \psi\vert^4$ is generalized to $\vert \psi\vert^{4+\varepsilon}$). Both models depend on a new parameter ($\varepsilon$) but when this parameter vanishes they become the familiar integrable models (the Sine-Gordon and NLS respectively). 

Our numerical results have shown that, in contradistinction to the integrable cases, the integrability-like constraints hold only for some field configurations and that they hold for the field configurations involving the scatterings of two solitons.
Hence we have tried to formulate the concept by expanding the models around their integrable points and specializing the discussion by looking at field configurations involving two solitons. Our results have then shown that some field configurations are very special; namely, when the field configurations possessed an extra parity-like symmetry  the models, for these configurations, possessed extra conserved quantities (very much like their integrable cousins). For other 
configurations these quantities were not conserved but were very small - with the meaning of `small' very dependent on the field configuration and on the closeness of the model to being integrable. Hence our views on quasi-integrability were very much tied up with these extra symmetries and this closeness.

To investigate this further we have decided to look at other topological models, in (1+1) dimensions; to check whether 
this relation of the quasi-integrability to the symmetry of the field configuration holds also in other models.

Hence we have decided to look at this in three topological models, dependent on a free parameter $\varepsilon$, which for a specific value of this parameter (namely $\varepsilon=0$) reduce to the Sine-Gordon model. 
The models were constructed using the procedure discussed in \cite{new},
which, in fact, is a generalization of the procedure discussed in \cite{bazeia}.
The first two models are obtained by two different changes of variables from the field of the Sine-Gordon model. In each case the one soliton field configuration connects the vacua at 0 and $\frac{\pi}{2}$. In one case there is a further vacuum at $-\frac{\pi}{2}$ irrespective of the value of the free parameter $\varepsilon$ while in the other one all other vacua depend on $\varepsilon$. Hence the first model allows for the symmetry of the two-soliton field configuration while the second one does not.  Finally, we have considered also a third
model, which depends on two parameters. When one of these two parameters is zero we have a situation  like in the first class
of models but when both parameters are nonzero we can study the transition from one regime to the other one and its effect 
on the properties of the anomaly.

The solitons reflect in their scatterings and so their effects on the anomalies are restricted only to the period when they are close together. However, we can also look at the configurations resembling breathers of the Sine-Gordon model.
Such configurations involve one soliton and one anti-soliton which are bound together (their energy is less than the energy 
of a soliton and an anti-soliton). Hence they are always in interaction and during this interaction they radiate some energy.
Thus the time evolution of their anomalies is stronger and so they can tell us more of the relevance of our symmetry
on quasi-integrability. Hence in the sequel we look at such configurations in the third class of models.

The paper is organized as follows. Section \ref{sec:model} presents our models, their properties and discusses their two soliton field configurations.
Section \ref{sec:analytical} recalls our ideas on quasi-integrability and their relation to the symmetry of the two-soliton configuration.
Our ideas are then tested, as described in section \ref{sec:numerical}, in numerical simulations of the scattering properties of two soliton configurations and time evolution of breather-like configurations.
We show that they support our expectations that the anomaly effects are smaller in models with symmetry and more
pronounced in the models in which this symmetry is broken.
We finish the paper with further comments and some conclusions, given in section \ref{sec:conclusions}.

\section{The models}
\label{sec:model}
\setcounter{equation}{0}

The three classes of models we consider in this paper are based on the Lagrangian density given by
\be
L\,=\,\frac{1}{2}\left((\partial_t \phi)^2\,-\,(\partial_x\phi)^2\right)\,-\,V(\phi),
\lab{one.one}
\ee
where the potentials $V(\phi)$ are conveniently chosen by the procedure discussed in \cite{new}, which we now briefly explain.
Following that procedure we take the field $\psi$ of the Sine-Gordon model, defined by the Lagrangian
\be
L_{\rm SG}\,=\,\frac{1}{2}\left((\partial_t \psi)^2\,-\,(\partial_x\psi)^2\right)\,-\,V_{\rm SG}\; ; \qquad\qquad \qquad V_{\rm SG}=\frac{1}{8}\,\sin^2(2\psi)
\lab{one.onea}
\ee
The static one-soliton solutions of the sine-Gordon model, given by 
\be
\psi\,=\,{\rm Arc Tan} (e^{\pm\(x-x_0\)}), 
\lab{one.two}
\ee
 are solutions of the so-called BPS equation
\be
\partial_x\psi=\pm \sqrt{2\, V_{\rm SG}}
\lab{bpssg}
\ee
Indeed, any solution of \rf{bpssg} is a solution of the static Euler-Lagrange equation associated to \rf{one.onea}.

Suppose now we introduce a field $\phi$ as a given transformation of the sine-Gordon field $\psi$, i.e. $\phi\equiv \phi\(\psi\)$. Then, it follows from \rf{bpssg} that
\be
\partial_x\phi=\pm \sqrt{2\, V\(\phi\)}
\lab{bpsphi}
\ee  
with
\be
V\(\phi\)\equiv \left(\frac{d\phi}{d\psi}\right)^2\,V_{\rm SG}=
\frac{1}{8}\sin^2\left[2\psi(\phi)\right]\, \left(\frac{d\phi}{d\psi}\right)^2
\lab{one.three}
\ee
If one takes the theory \rf{one.one} with potential given by \rf{one.three}, then it follows that static solutions of \rf{bpsphi} are solutions of the static Euler-Lagrange equation associated to \rf{one.one}. Therefore, the mapping $\phi\equiv \phi\(\psi\)$, maps BPS solutions of the sine-Gordon model into BPS solutions of the theory \rf{one.one}. 

In this paper we shall consider three theories of the type \rf{one.one}, with potentials given by  \rf{one.three}, for three particular choices of mappings $\phi\equiv \phi\(\psi\)$, which we now describe. 
\subsection{First class of models}

\begin{figure}[tbp]
    \centering
    \includegraphics[width=0.55\textwidth]{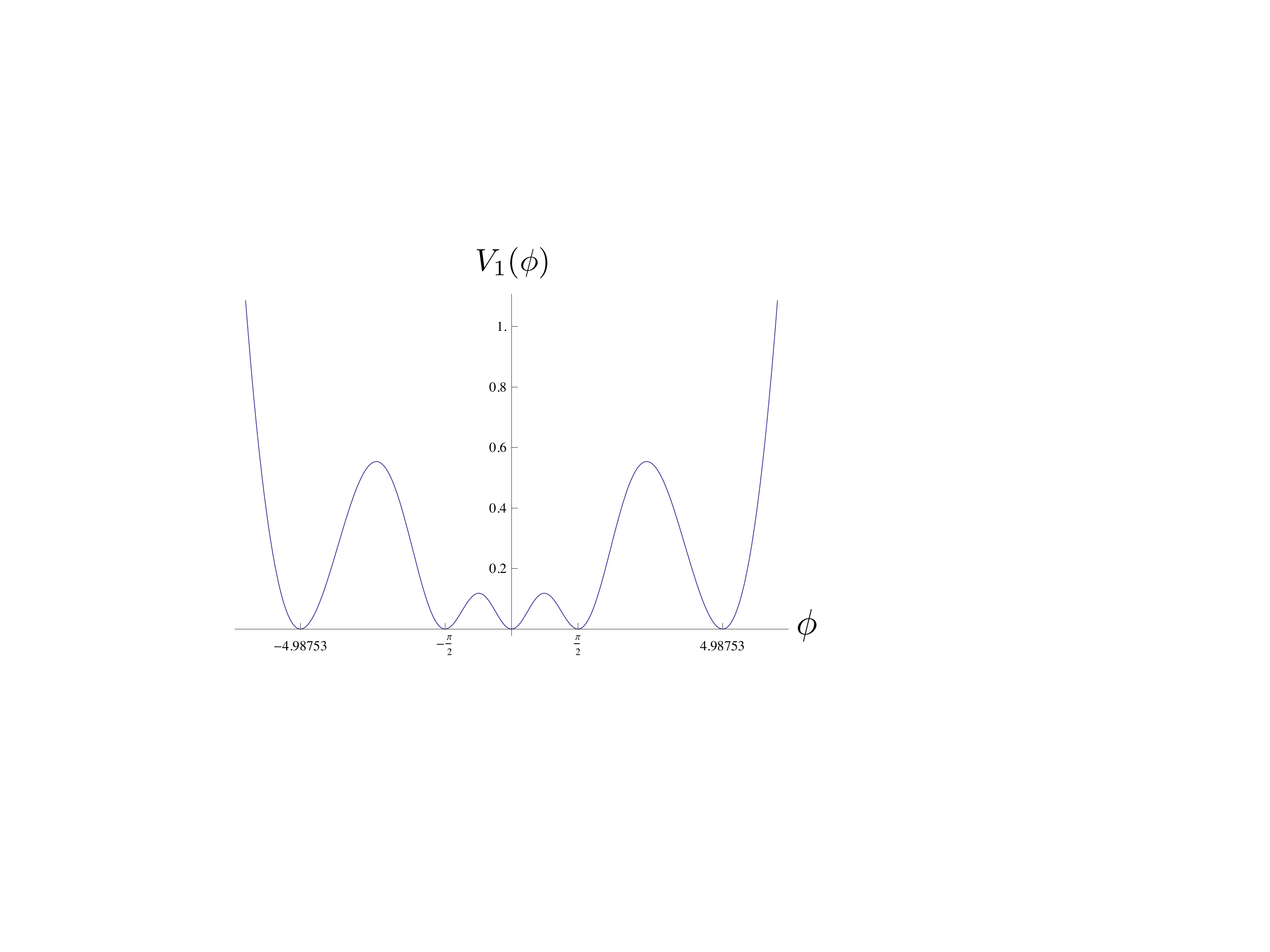}
    \begin{quote}
	\caption[AS]{\small Plot of the potential $V_1\(\phi\)$, given in \rf{v1def}, against $\phi$ for $\ve=0.05$. Note that  $V_1\(\phi\)$ is invariant under the change $\phi\rightarrow -\phi$, but its infinite number of vacua are not equally spaced. Therefore, a one-soliton solution tunneling between the vacua $\phi=0$ and $\phi=\frac{\pi}{2}$ have a mass which is different from, for instance, that of the one-soliton tunneling between the vacua $\phi=\frac{\pi}{2}$ and $\phi=4.98753$.} 
	\label{fig:plotv1}
    \end{quote}
\end{figure}

In the first class of models we perform the change of variables:
\be
\phi\(\psi\)\,=\,\frac{\psi}{1+\varepsilon(\frac{\pi^2}{4}-\psi^2)}.
\label{one.five}
\ee
where $\varepsilon$ is a free parameter which  has to satisfy
$\varepsilon>-\frac{4}{\pi^2}$ and when $\varepsilon=0$ we have the identity map, and so \rf{one.one} becomes the sine-Gordon model. Note that $\phi\(\psi=0\)=0$ and $\phi\(\psi=\pm\frac{\pi}{2}\)=\pm\frac{\pi}{2}$. Therefore, the three neighboring vacua of the sine-Gordon model, namely, $\psi=0,\pm\frac{\pi}{2}$, will  also be vacua of the new model. The same is not true however for the other SG vacua.  The inverse of this change of variables is provided by
\be
 \psi\,=\,\frac{2\,a\,\phi}{1+\Gamma}, \qquad\qquad{\rm with} \qquad \qquad  a=1+\frac{\varepsilon\,\pi^2}{4} \qquad\qquad \Gamma =\sqrt{1+4\,\varepsilon\,  a\,\phi^2}
\label{one.six}
\ee
One can easily check that
\be
\frac{d\psi}{d\phi}\,=\,\frac{2\,a}{\Gamma\,\(1+\Gamma\)}
\lab{one.eight}
\ee
Therefore, our first class of theory is defined by the Lagrangian \rf{one.one} with potential given by
 \be
V_1\(\phi\)=\frac{1}{32\, a^2}\, \Gamma^2\(1+\Gamma\)^2\,
\sin^2\left(\frac{4a\phi}{1+\Gamma}\right)
\lab{v1def}
\ee
Note that static one soliton solution of such a theory is given by (\ref{one.five}) in which $\psi$ is given by \rf{one.two}. We provide in figure \ref{fig:plotv1} a plot of the potential \rf{v1def} for $\ve=0.05$. 

\subsection{Second class of models}

\begin{figure}[tbp]
    \centering
    \includegraphics[width=0.6\textwidth]{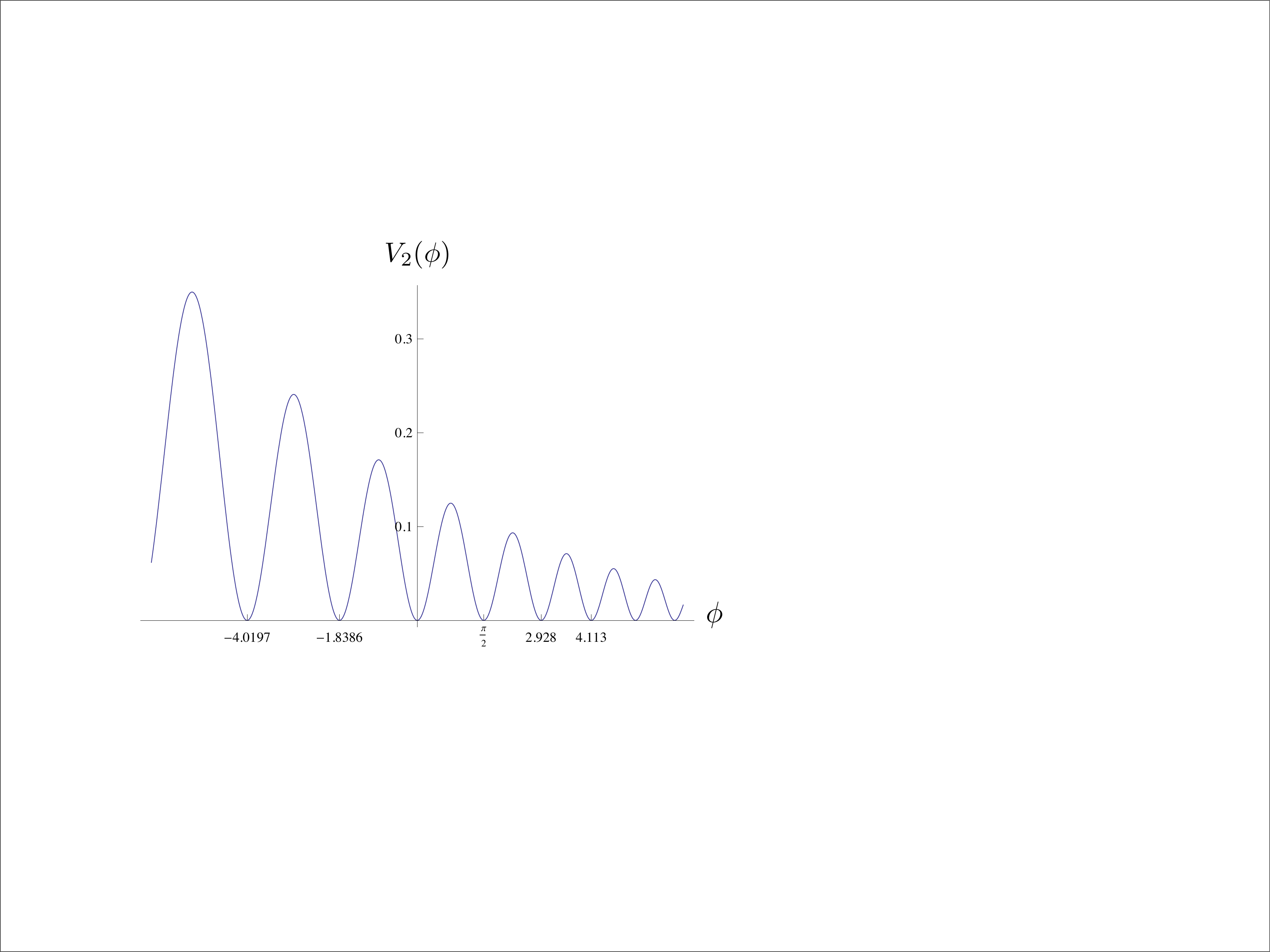}
    \begin{quote}
	\caption[AS]{\small Plot of the potential $V_2\(\phi\)$, given in \rf{v2def}, against $\phi$ for $\ve=0.05$. Contrary to $V_1$ this potential $V_2$ is not invariant under the change $\phi\rightarrow -\phi$, and its (infinite number) vacua are not equally spaced. Therefore, the one-soliton solutions tunneling between neighbouring vacua do not have the same mass.} 
	\label{fig:plotv2}
    \end{quote}
\end{figure}

In this case we make a change of variables:
\be \phi\,=\,\frac{b \psi}{1+\varepsilon \psi}\; ; \qquad\qquad 
 \psi\,=\,\frac{\phi}{(b-\varepsilon\phi)}\; ;\qquad\qquad {\rm with}\qquad
 b=1+\varepsilon\frac{\pi}{2}
\lab{one.ninea}
\ee
where the choice of the parameter $b$ such that  $\phi\(\psi=0\)=0$ and $\phi\(\psi=\frac{\pi}{2}\)=\frac{\pi}{2}$. Therefore, one has
\be
\frac{d\phi}{d\psi}\,=\,\frac{(b-\varepsilon\phi)^2}{b}.
\lab{one.eleven}
\ee
and so our second class of theories is defined by the Lagrangian  \rf{one.one} with the potential being
\be
V_2\(\phi\)=\frac{1}{8}
\frac{(b-\varepsilon\phi)^4}{b^2}\,\sin^2\left(\frac{2\phi}{b-\varepsilon\phi}\right)
\lab{v2def}
\ee
Again the static one soliton solution of this model is given by (\ref{one.five}) in which $\psi$ is given by \rf{one.ninea}. We give in figure \ref{fig:plotv2} a plot of the potential \rf{v2def} for $\ve=0.05$. 

\subsection{Third class of Models}

Given the very different behaviour of both models presented above, and given the numerical results which we obtained for them which will be discussed in section \ref{sec:numerical}, 
we have decided to look also at a model which interpolates between the two of them.
 To study this in more detail we have decided to construct a  third model (with two parameters). This would allows us to study the transition from 
one behaviour to the other  as we change the parameters. 
We then introduce the third class of models based on the change of variables
\be
\psi\(\phi\)\,=\,\frac{c\,\phi}{\sqrt{1+\varepsilon\,\phi\,(\phi-2\,\gamma)}}.
\lab{three.one}
\ee
where
\be
c\,=\,\sqrt{1+\varepsilon\,\pi\left(\frac{\pi}{4}-\gamma\right)},
\lab{three.two}
\ee
The parameter $c$ was chosen such that $\phi\(\psi=0\)=0$ and $\phi\(\psi=\frac{\pi}{2}\)=\frac{\pi}{2}$. The free parameters are $\ve$ and $\gamma$.

It then follows that the third class of theories is defined by the Lagrangian \rf{one.one} with the potential being given by
\be
V_3(\phi)\,=\,\frac{1}{8} \,\frac{(1+\varepsilon\phi(\phi-2\gamma))^3}{c^2\,(1-\varepsilon\gamma\phi)^2}\,\sin^2(2\, \psi(\phi)),
\lab{v3def}
\ee
where $\psi\(\phi\)$ is given by \rf{three.one}. Note that  when $\gamma=0$ the model possesses the  symmetry $\phi \rightarrow -\phi$, which will be desirable for quasi-integrability as we discuss it further in the paper. Then the $\gamma =0$ case  resembles a little the first class models given by the potential \rf{v1def}. The case where $\gamma\ne0$ we do not have any symmetry and so the model is more like the models of the second class, given by the potential \rf{v2def}.  When $\varepsilon=0$, the parameter $\gamma$ becomes irrelevant and we have the sine-Gordon model. Consequently,  this third class of models is a good place to study the transition mentioned above.

\begin{figure}[tbp]
    \centering
    \includegraphics[width=1\textwidth]{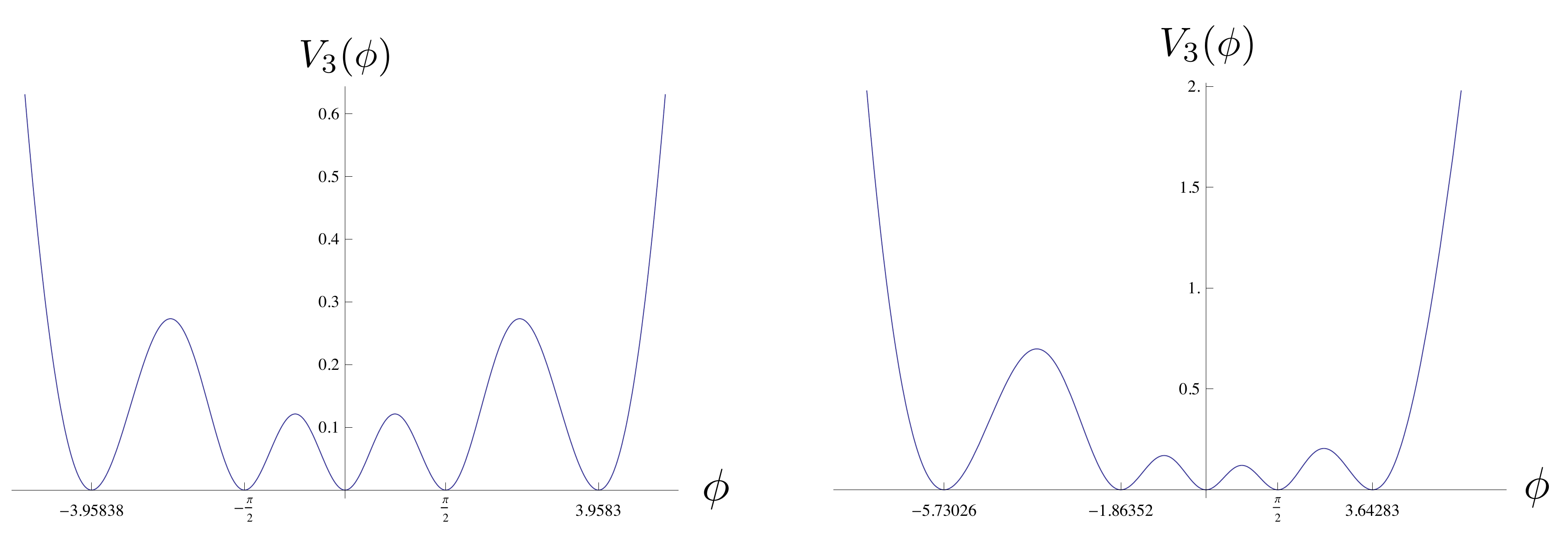}
    \begin{quote}
	\caption[AS]{\small Plots of the potential $V_3\(\phi\)$, given in \rf{v3def}, against $\phi$ for $\ve=0.05$, with $\gamma=0$ for the left plot, and $\gamma=1$ for the right plot. Note that this class of potentials interpolates between the first class (for $\gamma=0$) and the second class (for $\gamma\neq 0$). Note that, for $\gamma=0$, $V_3$ is symmetric under the change $\phi\rightarrow -\phi$, but not for other values of $\gamma$, and as long as $\ve\neq 0$, the infinite number of vacua are not equally spaced, leading to one-soliton solutions of different masses.} 
	\label{fig:plotv30}
    \end{quote}
\end{figure}

\section{Quasi-integrability}
\label{sec:analytical}
\setcounter{equation}{0}

Next we perform an analytical study of the properties of our models using the techniques usually employed in integral field theories to construct quantities which are quasi-conserved as will be explained below. We follow the ideas put forward in  our previous works on quasi-integrability  \cite{us,us2}. 

We consider a scalar field theory for a real scalar field $\phi$ in a generic potential $V\(\phi\)$. As in \cite{us} we introduce the Lax potentials, based on a $sl(2)$ loop algebra:
\br
A_{+}&=& \frac{1}{2}\left[ \(\omega^2 \, V -m\)\, b_{1}
  -i\,\omega\, \frac{d\,V}{d\,\phi}\,F_1\right], 
\nonumber\\
A_{-}&=& \frac{1}{2}\, b_{-1} - \frac{i}{2}\,
\omega\, \partial_{-}\phi\, F_0, 
\lab{potentials}
\er
where we have used light-cone variables 
\be
x_{\pm}=\frac{1}{2}\(t\pm x\) \qquad {\rm with} \qquad  \partial_{\pm}=\partial_t\pm \partial_x \qquad {\rm and} \qquad  \partial_{+}\partial_{-}=\partial_t^2-\partial_x^2\equiv \partial^2.
\lab{lightconedef}
\ee
The $sl(2)$ loop algebra generators $b_n$ and $F_n$ are defined in the appendix \ref{sec:appendix-algebra}, and $\omega$ and $m$ are free parameters which do not appear in the Lagrangian or the equations of motion of the theory. One can easily check that the curvature of the connection \rf{potentials} is given by
\be
F_{+-}\equiv \partial_{+}A_{-}-\partial_{-}A_{+}+\sbr{A_{+}}{A_{-}}= X
\, F_1 -\frac{i\,\omega}{2}\left[\partial^2 \phi + \frac{\partial\,
    V}{\partial\, \phi} \right]\,F_0
\lab{zc}
\ee
with
\be
X = \frac{i\,\omega}{2}\,  \partial_{-}\phi\,
\left[\frac{d^2\,V}{d\,\phi^2}+\omega^2\, V-m\right].
\lab{xdef}
\ee

Note that when the equations of motion, following from \rf{one.one}, are satisfied, {\it i.e.} 
\be
\partial^2 \phi + \frac{\partial\,V}{\partial\, \phi}=0
\lab{eqofmot}
\ee
the term proportional to the Lie algebra generator $F_0$ vanishes. In addition, the quantity $X$ vanishes for the sine-Gordon potential
\be
V_{\rm SG}=\frac{1}{16}\, \left[1-\cos\(4\,\phi\)\right]
\lab{sgpot}
\ee
if one chooses the parameters $m$ and $\omega$  as $m=1$ and $\omega=4$. Therefore, the curvature vanishes for the sine-Gordon theory, and that is what makes this theory an integrable field theory in $(1+1)$ dimensions. For any other choice of the potential $V\(\phi\)$, however, the curvature does not vanish. Following the steps given in \cite{us} now analyze what can be said about conservation laws in a theory with non-vanishing $X$, or equivalently $F_{+-}$.
   
We start by recalling the fact that the generator $b_{-1}$ is a semi-simple element in the sense that its adjoint action splits the $sl(2)$ loop algebra ${\cal G}$ into its kernel and image as follows:
\be
{\cal G}= {\rm Ker} + {\rm Im}, \qquad\qquad \sbr{b_{-1}}{{\rm Ker}}=0,\qquad \qquad{\rm Im}=\sbr{b_{-1}}{{\cal G}}
\ee
with the important property  that ${\rm Ker}$ and ${\rm Im}$ have no common elements. An important ingredient of the method is that our $sl(2)$ loop algebra
  ${\cal G}$ is graded by the grading operator $d= T_3+
  2\,\lambda \frac{d\;}{d\lambda}$ (see appendix
  \ref{sec:appendix-algebra} for details),  with the integer grades $n$
  being  determined   as 
\be
{\cal G}=\sum_{n}{\cal G}_n\; ; \qquad\qquad \sbr{{\cal G}_m}{{\cal
    G}_n}\subset {\cal G}_{m+n}\; ; \qquad\qquad \sbr{d}{{\cal
    G}_n}= n \, {\cal G}_n.
\ee

Next we perform a gauge transformation 
\be
A_{\mu}\rightarrow a_{\mu}=g\, A_{\mu}\,g^{-1}-\partial_{\mu}g\,
g^{-1}
\lab{gaugeminus}
\ee
with the group element $g$ being an exponentiation of generators lying in
the positive grade subspace generated by the $F_n$'s, {\it i.e.}, 
\be
g={\rm exp}\left[\sum_{n=1}^{\infty} {\cal F}^{(n)}\right] \qquad\qquad \qquad {\rm with} \qquad\qquad {\cal F}^{(n)}=\zeta_n\, F_n
\lab{gplusdef}
\ee
and $\zeta_n$, being the parameters of the transformation, which will be determined as we will
explain below. Under \rf{gaugeminus} the curvature \rf{zc}  transforms as
\be
F_{+-}\rightarrow
g\,F_{+-}\,g^{-1}=\partial_{+}a_{-}-\partial_{-}a_{+}+\sbr{a_{+}}{a_{-}}=
X \, g\, F_1\,g^{-1},
\lab{newcurvature}
\ee
where we have used the equations of motion \rf{eqofmot} to drop the term
proportional to $F_0$ in \rf{zc}.

 The component $A_{-}$ of the
connection \rf{potentials} has terms with grades $0$ and
$-1$. Therefore, under \rf{gaugeminus} it gets transformed into
$a_{-}$ which has terms with grades ranging from $-1$ to
$+\infty$. Decomposing $a_{-}$ into grades we get from \rf{gaugeminus}
and \rf{gplusdef} that
\br
a_{-}&=&\frac{1}{2}\,b_{-1}
\lab{determinezeta}\\
&-&\frac{1}{2}\,\sbr{b_{-1}}{{\cal F}^{(1)}}- \frac{i}{2}\,
\omega\, \partial_{-}\phi\, F_0\nonumber\\
&-&\frac{1}{2}\,\sbr{b_{-1}}{{\cal F}^{(2)}}
+\frac{1}{4}\,\sbr{\sbr{b_{-1}}{{\cal F}^{(1)}}}{{\cal F}^{(1)}}
- \frac{i}{2}\,
\omega\, \partial_{-}\phi\, \sbr{{\cal F}^{(1)}}{F_0}
-\partial_{-}{\cal F}^{(1)}
\nonumber\\
&\vdots&\nonumber\\
&-&\frac{1}{2}\,\sbr{b_{-1}}{{\cal F}^{(n)}}+ \ldots. \nonumber
\er

Next we note that we can make the component of $a_{-}$ in the direction of $F_{n-1}$ vanish by choosing the parameters $\zeta_n$.
In fact we can do this recursively. In particular,  
$\zeta_1= \frac{i}{2}\,
\omega\, \partial_{-}\phi$, and so on.  Thus our gauge transformation \rf{gplusdef} has rotated  the component $a_{-}$ into the abelian
subalgebra generated by the $b_{2n+1}$'s.
 Note that in this procedure we 
have not used the equations of motion \rf{eqofmot}.
 Thus
\be
a_{-}=\frac{1}{2}\,b_{-1}+\sum_{n=0}^{\infty}a_{-}^{(2n+1)}\, b_{2n+1}
\ee

Next we note that, even when the equations of motion \rf{eqofmot} are used, the component 
$a_{+}$ of the connection does not get rotated into the abelian subalgebra generated by the $b_{2n+1}$'s ({\it i.e.} the kernel), as it 
 has components in the image too, {\it i.e.} it is of the form: 
\be
a_{+}=\sum_{n=0}^{\infty}a_{+}^{(2n+1)}\, b_{2n+1}
+\sum_{n=2}^{\infty} c_{+}^{(n)}\,F_n.
\ee
It turns out that the coefficients $c_{+}^{(n)}$ are all proportional to $X$, given in \rf{xdef}, and so they vanish for the sine-Gordon theory. As we shall see below this vanishing of $X$ is what is responsible for the integrability of the sine-Gordon theory as it possesses an infinite set of exact conservation laws. 

The next step of our procedure, as discussed in \cite{us} involves the decomposition of the curvature \rf{newcurvature} into two 
components, one lying in the abelian subalgebra generated by $b_{2n+1}$'s  and the other lying in the subspace generated by $F_n$'s. 
Thus, 
\be
g\,F_1\,g^{-1}=\sum_{n=0}^{\infty} \gamma^{(2n+1)}\, b_{2n+1} + \mbox{\rm terms proportional to $F_n$'s}
\lab{f1conjugate}
\ee

Since $a_{+}$ lies in the kernel it follows that the commutator $\sbr{a_{+}}{a_{-}}$ does not produce terms in the kernel. Therefore, the terms in the direction of the $b_{2n+1}$'s of the transformed curvature \rf{newcurvature}  lead to
 the equations of the form
\be
\partial_{+}a_{-}^{(2n+1)}-\partial_{-}a_{+}^{(2n+1)}=
X\,\gamma^{(2n+1)}\qquad\qquad n=0,1,2,\ldots
\lab{quasiconserv}
\ee
with  the anomaly $X$ given in \rf{xdef}.

If we rewrite this using $x$ and $t$ variables we find  that \rf{quasiconserv}
takes the form $\partial_{t}a_{x}^{(2n+1)}-\partial_{x}a_{t}^{(2n+1)}=-\frac{1}{2}\,
X\,\gamma^{(2n+1)}$, and so we find that (see \rf{lightconedef}) 
\be
\frac{d\,Q^{(2n+1)}}{d\,t}=-\frac{1}{2}\,\alpha^{(2n+1)}+
a_{t}^{(2n+1)}\mid_{x=-\infty}^{x=\infty}, 
\lab{timedercharges}
\ee
where
\be
Q^{(2n+1)}\equiv \int_{-\infty}^{\infty}dx\,a_{x}^{(2n+1)},\qquad\qquad\qquad
\alpha^{(2n+1)}\equiv \int_{-\infty}^{\infty}dx\,X\,\gamma^{(2n+1)}.
\lab{chargeanomalydef}
\ee

For finite energy solutions the fields $\phi$ have to go to a vacuum configuration at spatial infinity and, in general, this condition is enough for the values of $a_{t}^{(2n+1)}$ at $x=\pm \infty$ to be equal. Therefore, the non-conservation of the charges is due to the non-vanishing of the anomalies $\alpha^{(2n+1)}$, and so of the non-vanishing of the quantity $X$ given in \rf{xdef}.

To proceed further we introduce the following order two automorphism of the $sl(2)$ loop algebra 
\be
\Sigma\(b_{2n+1}\)=- b_{2n+1},\qquad\qquad \Sigma\(F_{2n}\)=-F_{2n}, \qquad\qquad \Sigma\(F_{2n+1}\)=F_{2n+1}
\lab{automorphism}
\ee 
which is, in fact,  the same as 
\be
\Sigma\(T_3\)=-T_3,\qquad\qquad \Sigma\(T_{\pm}\)=-T_{\mp}
\ee
since the so-called spectral parameter $\lambda$ of the loop algebra  is not affected by $\Sigma$ (see the appendix \ref{sec:appendix-algebra}). 
At the same time we introduce the space-time parity transformation
\be
P:\qquad \({\tilde x},{\tilde t}\)\rightarrow \(-{\tilde x},-{\tilde t}\) \qquad\qquad {\rm with} \qquad \quad{\tilde x}= x-x_{\Delta} \qquad \quad{\tilde t}=t-t_{\Delta},
\lab{paritydef}
\ee
where $x_{\Delta}$ and $t_{\Delta}$ are constants which, as we will see later, are  determined by the particular solution under consideration. Even though the values of these constants vary from a solution to a solution, the argument we are presenting is independent of their values. 

Next we make an important hypothesis that will have to be verified for each particular solution of the equations of motion that we want to study. We assume that, under the parity $P$, the scalar field $\phi$, evaluated on the solution under consideration, transforms as
\be
\phi \rightarrow -\phi+{\rm const.}
\lab{niceparitytransf}
\ee
Thus,  from \rf{potentials},  we see that
\be
\Sigma\( A_{-}\)=-A_{-} \qquad\qquad \qquad P\(A_{-}\)=A_{-}
\ee
and so
\be
\Omega\(A_{-}\)=-A_{-}\qquad\qquad\qquad {\rm with} \qquad\qquad \Omega\equiv \Sigma\, P.
\ee

Next we take the even part of \rf{determinezeta} under $\Omega$ and split it in its grade components. From the first line of \rf{determinezeta} we find that
\be
\(1+\Omega\)a_{-}\mid_{-1}= 0.
\ee
The second line of \rf{determinezeta} then implies that 
\be
\(1+\Omega\)a_{-}\mid_{0}= -\frac{1}{2}\,\sbr{b_{-1}}{\(1-\Omega\){\cal F}^{(1)}}.
\lab{zerocompomega}
\ee
Note that, since we have rotated $a_{-}$ into  the ${\rm Ker}$, the l.h.s. of \rf{zerocompomega} has to lie in this ${\rm Ker}$.
 However, the r.h.s. of \rf{zerocompomega} lies  in the ${\rm Im}$. Therefore both sides have to vanish, {\it i.e.} 
\be
\(1+\Omega\)a_{-}\mid_{0}=0,\qquad\qquad\qquad \(1-\Omega\){\cal F}^{(1)}=0.
\ee
Note also, using the fact that ${\cal F}^{(1)}$ is even under $\Omega$, that the third line of \rf{determinezeta} gives
\be
\(1+\Omega\)a_{-}\mid_{1}= -\frac{1}{2}\,\sbr{b_{-1}}{\(1-\Omega\){\cal F}^{(2)}}
\ee
and so using the same arguments as above we see that 
\be
\(1+\Omega\)a_{-}\mid_{1}=0,\qquad\qquad\qquad \(1-\Omega\){\cal F}^{(2)}=0.
\ee

These arguments can now be repeated and so wee that all ${\cal F}^{(n)}$ are even under $\Omega$ and so, in consequence,  the group element performing the gauge transformation \rf{gaugeminus} satisfies
\be
\Omega\(g\)=g.
\ee

The Killing form of the $sl(2)$ loop algebra is given by 
\be
{\rm Tr}\(b_{2m+1}\,b_{2n+1}\)= c \,\delta_{2\(m+n\)+2,0},\qquad\qquad \qquad
{\rm Tr}\(b_{2m+1}\,F_n\)=0
\lab{killingform}
\ee
for some constant $c$. Thus, using \rf{killingform}, we find from \rf{f1conjugate} that 
\be
\gamma^{(2n+1)}=\frac{1}{c}\, {\rm Tr}\(g\,F_1\,g^{-1}\, b_{-2n-1}\) 
=-\frac{1}{c}\, {\rm Tr}\(\Sigma\(g\)\,F_1\,\Sigma\(g^{-1}\)\, b_{-2n-1}\) 
\ee
and so
\be
P\(\gamma^{(2n+1)}\)=-\frac{1}{c}\, {\rm Tr}\(\Omega\(g\)\,F_1\,\Omega\(g^{-1}\)\, b_{-2n-1}\) 
=-\frac{1}{c}\, {\rm Tr}\(g\,F_1\,g^{-1}\, b_{-2n-1}\) =-\gamma^{(2n+1)}.
\ee

Next we make our second hypothesis that will have to be verified for each particular potential; namely, that the quantity $X$
 defined in \rf{xdef} satisfies 
\be
P\(X\)=X.
\lab{parityx}
\ee
This can be achieved, for example, if the potential $V$, evaluated on the solution, is even under the parity $P$ given in \rf{paritydef}. If this is  true it then follows that 
\be
\int_{-{\tilde t}_0}^{{\tilde t}_0} dt\,\int_{-{\tilde x}_0}^{{\tilde x}_0} dx\, X\, \gamma^{(2n+1)}=0,
\lab{mm}
\ee
where ${\tilde t}_0$ and ${\tilde x}_0$ are any given fixed values of the shifted time ${\tilde t}$ and space coordinate ${\tilde x}$, respectively, introduced in \rf{paritydef}.
It this is the case thenh,  by taking ${\tilde x}_0\rightarrow \infty$, we conclude that the non-conserved charges \rf{chargeanomalydef} satisfy the following mirror time-symmetry around the point: $t_{\Delta}$.
\be
Q^{(2n+1)}\(t={\tilde t}_0 +t_{\Delta}\)=Q^{(2n+1)}\(t=-{\tilde t}_0 +t_{\Delta}\) 
\qquad\qquad\qquad n=0,1,2,\ldots
\lab{mirrorcharge}
\ee

Another infinite set of quasi-conserved charges, satisfying \rf{mirrorcharge}, can be constructed by a connection like that given in \rf{potentials} but with $x_{+}$ and $x_{-}$ exchanging roles, and the grades changing sign. Then one performs  a gauge transformation  like in \rf{gaugeminus} but with the group element $g$ being an exponentiation like in \rf{gplusdef} but involving negative grade generators $F_{-n}$. See \cite{us} for the details of that construction. 

Summarizing, we conclude that
\begin{enumerate}
\item If we have a two-soliton-like or a breather-like solution of the equations of motion \rf{eqofmot}, transforming under the space-time parity \rf{paritydef} as in \rf{niceparitytransf}, {\it i.e.}
\be
P\(\phi\)=  -\phi+{\rm const.}
\lab{require1}
\ee
\item And if the potential, evaluated on such a solution is even under the parity, {\it i.e.}
\be
P\(V\)=V
\lab{require2}
\ee
such that \rf{parityx} holds true,
\item  Then we have an infinite set of conserved quantities which are conserved asymptotically, {\it i.e.}
\be
Q^{(2n+1)}\(t=+\infty\)=Q^{(2n+1)}\(t=-\infty\).
\lab{quasi}
\ee
So, the values of these charges in the infinite past, before the scattering of the solitons, are the same as in the infinite future, after the scattering.  For the  breather solutions such charges are symmetric by reflection around $t=t_{\Delta}$ (see \rf{mirrorcharge}), and in fact oscillate in time. Theories possessing such properties we call {\em quasi-integrable} theories.
\end{enumerate}

\subsection{Expansion around sine-Gordon}

Here we analyze under what circumstances  one can satisfy the conditions given above and so have {\em quasi-integrability}. We by start looking at the properties of the two-soliton and breather solutions of the sine-Gordon model. For the potential \rf{sgpot} the equation of motion \rf{eqofmot} becomes the sine-Gordon equation 
\be
\partial^2\phi_0+\frac{1}{4}\,\sin\(4\,\phi_0\)=0.
\lab{eqforphi0}
\ee

\subsubsection{The two-soliton solutions}

The two-soliton solution of \rf{eqforphi0} is given by 
\be
\phi_0= {\rm ArcTan}\, \frac{{\rm Im}\,\tau_0}{{\rm Re}\,\tau_0},
\lab{vp0tau0}
\ee
where the
$\tau_0$ function is given by
\br
\tau_0 =1 + i\, e^{\Gamma_1}+i\, e^{\Gamma_2}-
\gamma \, e^{\Gamma_1+\Gamma_2}
\lab{2solitontau}
\er
with
\be
\Gamma_i =
\eta_i\,\frac{\(x-v_i\,t-x_0^{(i)}\)}{\sqrt{1-v_i^2}}
\qquad\qquad \qquad i=1,2 
\lab{gammaidef}
\ee
and 
\be
\gamma=\(\tanh\(\frac{\alpha_2-\alpha_1}{2}\)\)^{2\ve_1\ve_2}
\lab{gammadef}
\ee
and where $\eta_i=\pm 1$, for kink and anti-kink. Here,
$v_i=\tanh \alpha_i$, is the velocity (in units of the speed of light)
of the kink (anti-kink) $i$. 

Note that we can write the solution \rf{vp0tau0} as 
\be
\phi_0={\rm ArcTan}\, \(-\frac{1}{\sigma}\),
\ee
where
\be
\sigma\equiv
\sqrt{\gamma}\,\frac{\sinh z_{+}}{\cosh z_{-}}
\ee
and
\be
z_{+}=\frac{\Gamma_1+\Gamma_2}{2} +\delta, \qquad \qquad \qquad
z_{-}=\frac{\Gamma_1-\Gamma_2}{2}, \qquad\qquad\qquad \delta=\ln \sqrt{\gamma}.
\ee

Note also that as $z_{+}$ and $z_{-}$ are linear functions of $x$ and $t$ they can be taken as new coordinates of the space-time.  They are,
of course, independent of each other due to the fact that in a two-soliton solution we can never put $v_1=v_2$. Indeed, if that happens the two-soliton solution reduces to a one-soliton solution for the case $\eta_1\,\eta_2=1$, and one gets a divergent factor ($\gamma\rightarrow \infty$), for $\eta_1\,\eta_2=-1$. In addition, we see from \rf{vp0tau0} that
\be
\sin \phi_0= \frac{{\rm Im}\,\tau_0}{\mid \tau_0\mid}= \frac{1}
{\sqrt{1+\sigma^2}},\qquad\qquad\qquad
\cos \phi_0= \frac{{\rm Re}\,\tau_0}{\mid \tau_0\mid}=-\frac{\sigma}.
{\sqrt{1+\sigma^2}}
\lab{sincosvp0}
\ee

Thus, under the space-time parity transformation
\be
P:\qquad \(z_{+},z_{-}\)\rightarrow \(-z_{+},-z_{-}\)
\lab{paritydefz}
\ee
we see that $P\(\sigma\)=-\sigma$, and so, taking the domain of ${\rm ArcTan}$
to be $\(0,\pi\)$, we have that 
\be
P\(\phi_0\)=-\phi_0+\pi.
\lab{parityphi0sg}
\ee
In terms of $x$ and $t$ the parity transformation \rf{paritydefz} takes the form of \rf{paritydef} with 
\br
x_{\Delta}&=&\frac{\eta_1\,\eta_2}{\sinh\(\alpha_1-\alpha_2\)}\,\left[
\eta_2\, \sinh \alpha_2\, (\delta-\eta_1 \,x_0^{(1)}\, \cosh \alpha_1 )-\eta_1\,
   \sinh \alpha_1\, (\delta -\eta_2 \,x_0^{(2)} \cosh \alpha_2)\right],
\nonumber\\
t_{\Delta}&=&\frac{\eta_1\,\eta_2}{\sinh\(\alpha_1-\alpha_2\)}\,\left[
\delta \, \eta_2 \,\cosh \alpha_2-\eta_1 \cosh \alpha_1\, (\eta_2\, (x_0^{(1)}-x_0^{(2)})\,\cosh
   \alpha_2 +\delta )\right].
\er

From \rf{parityphi0sg} we note that the  even and odd components of $\phi_0$ are given by
\be
\phi_0^{(\pm)}\equiv \frac{1}{2}\(1\pm P\)\(\phi_0\), \qquad \qquad \mbox{\rm and so}\qquad \qquad \phi_0^{(-)}=\phi_0-\frac{\pi}{2}\; ;\qquad\quad 
\phi_0^{(+)}=\frac{\pi}{2}
\ee
Note also that $\sin\(4\,\phi_0\)=\sin\(\phi_0^{(-)}\)$ and so we see that  $\phi_0^{(-)}$ and $\phi_0^{(+)}$ are both solutions of the sine-Gordon equation, {\it i.e.} 
\be
\partial^2\phi_0^{(\pm)}+\frac{1}{4}\,\sin\(4\,\phi_0^{(\pm)}\)=0.
\lab{eqforphi0minus}
\ee
In fact, $\phi_0^{(+)}$ is a vacuum solution and $\phi_0^{(-)}$ is a proper two-soliton solution (with $\phi_0$ being given by \rf{vp0tau0}). 

\subsubsection{The breather solutions}

The moving breather solution of the sine-Gordon equation \rf{eqforphi0} is also obtained from a tau function, thorough \rf{vp0tau0}, given by \cite{massformula}
\be
\tau_0 = 1+e^{2\,\Gamma_R} + 2\,i\,\({\rm cotan}\, \theta\)\;
e^{\Gamma_R}\, \sin \Gamma_I,
\ee
with 
\be
\Gamma_R = \frac{\cos\theta}{\sqrt{1-v^2}}\, \(x-v\,t\),\qquad\qquad 
\Gamma_I = \frac{\sin\theta}{\sqrt{1-v^2}}\, \(t-v\,x\).
\ee

So, from \rf{vp0tau0} one gets that
\be
\phi_0 =  {\rm ArcTan}\frac{\({\rm cotan}\, \theta\)\,\sin
  \Gamma_I}{\cosh \Gamma_R},
  \lab{breathersol}
\ee
where $v$ is the speed of the breather, and $\sin\theta$ is its frequency (in its rest frame). 
Note that the argument of ${\rm ArcTan}$ in \rf{breathersol} never diverges and so $\phi_0$ never passes through $\pm \frac{\pi}{2}$. Thus, for the breather solution it is more convenient to take the domain of ${\rm ArcTan}$ to be $\(-\frac{\pi}{2},\frac{\pi}{2}\)$. We now introduce the space-time parity transformation
\be
P\; : \qquad\qquad \(x\, , \, t\)\rightarrow  \(-x\, , \, -t\)
\ee
and comparing it with \rf{paritydef} we see that in this case $x_{\Delta}=t_{\Delta}=0$. Under this parity transformation, the breather solution \rf{breathersol} is odd, {\it i.e.} 
\be
P\(\phi_0\)=-\phi_0
\lab{breatherparity}
\ee
So, in the breather case we see that
\be
\phi_0^{(\pm)}\equiv \frac{1}{2}\(1\pm P\)\(\phi_0\) \qquad \qquad \rightarrow \qquad \qquad \phi_0^{(-)}=\phi_0\; ;\qquad\quad 
\phi_0^{(+)}=0
\ee
and so $\phi_0^{(-)}$ and $\phi_0^{(+)}$ are again both solutions of the sine-Gordon equation \rf{eqforphi0}, with the first being the breather itself and the second a trivial vacuum solution.

\subsubsection{The parity versus dynamics argument for our models}

We start by observing that the sine-Gordon potential \rf{sgpot} is invariant under \rf{parityphi0sg} and \rf{breatherparity}. Therefore, the sine-Gordon theory satisfies the requirements \rf{require1} and \rf{require2}, for quasi-integrability, within the two-soliton and breather sector, as it should, since it is an exact integrable field theory.  We now discuss the conditions under which the theories, which are smooth deformations of the sine-Gordon, should satisfy to be called quasi-integrable. 

We shall restrict our attentions to the theories of the previous section; {\it i.e.} to theories with equations of motion given by \rf{eqofmot}, where the potential $V$ is a smooth deformation of the sine-Gordon potential \rf{sgpot}, in the sense that $V$ depends upon a parameter $\ve$ such that for $\ve=0$, $V$ becomes \rf{sgpot}. In order to perform the expansion around the sine-Gordon model we shall work with  two-soliton and breather solutions which are odd under the parity. We will also expand the field $\phi$, as a solution of \rf{eqofmot},  in a power series in $\ve$, {\it i.e.}
\be
\phi=\phi_0^{(-)}+\ve\, \phi_1+\ve^2\,\phi_2+\ldots.
\ee

Thus, the potential $V$ depends upon $\ve$ explicitly, and also implicitly through $\phi$. Then the Taylor expansion around $\ve=0$ of its first derivative w.r.t. $\phi$ takes the form 
\br
\frac{\partial\,V}{\partial\,\phi} &=& 
\frac{\partial\,V}{\partial\,\phi} \mid_{\ve=0} +
\left[\frac{d\,}{d\,\ve}\(\frac{\partial\,V}{\partial\,\phi}
  \)\right]_{\ve=0}\, \ve +\ldots\nonumber\\
&=& 
\frac{\partial\,V}{\partial\,\phi} \mid_{\ve=0} +
 \left[\frac{\partial^2 V}{\partial \ve\partial\phi}+
\frac{\partial^2 V}{\partial \phi^2}\,\frac{\partial \phi}{\partial
  \ve}\right]_{\ve=0} \, \ve \nonumber\\
&+&\frac{1}{2}\,
 \left[\frac{\partial^3 V}{\partial \ve^2\partial\phi}
+
2\,\frac{\partial^3 V}{\partial\ve\partial \phi^2}\,\frac{\partial \phi}{\partial
  \ve}
+
\frac{\partial^2 V}{\partial \phi^2}\,\frac{\partial^2 \phi}{\partial
  \ve^2}
+
\frac{\partial^3 V}{\partial \phi^3}\,\(\frac{\partial \phi}{\partial
  \ve}\)^2\right]_{\ve=0} \, \ve^2 
+\ldots
\er

In consequence, splitting \rf{eqofmot} in powers of $\ve$ we find that $\phi_0$ satisfies the sine-Gordon equation \rf {eqforphi0}. The higher components of the field $\phi$ satisfy
\br
\partial^2 \phi_1+ \frac{\partial^2 V}{\partial \phi^2}\mid_{\ve=0}\, \phi_1&=&
-\frac{\partial^2 V}{\partial \ve\partial\phi}\mid_{\ve=0},
\lab{phi1eq}\\
\partial^2 \phi_2+ \frac{\partial^2 V}{\partial \phi^2}\mid_{\ve=0}\, \phi_2&=&
-\frac{1}{2}\,
 \left[\frac{\partial^3 V}{\partial \ve^2\partial\phi}\mid_{\ve=0}
+
2\,\frac{\partial^3 V}{\partial\ve\partial \phi^2}\mid_{\ve=0}\,\phi_1
+
\frac{\partial^3 V}{\partial \phi^3}\mid_{\ve=0}\,\phi_1^2\right], 
\lab{phi2eq}\\
\partial^2 \phi_3+ \frac{\partial^2 V}{\partial \phi^2}\mid_{\ve=0}\, \phi_3&=&
-\frac{1}{3!}\,
 \left[\frac{\partial^4 V}{\partial \ve^3\partial\phi}\mid_{\ve=0}
+
3\,\frac{\partial^4 V}{\partial\ve^2\partial \phi^2}\mid_{\ve=0}\,\phi_1
+
6\,\frac{\partial^3 V}{\partial\ve\partial \phi^2}\mid_{\ve=0}\,\phi_2
\right. \nonumber\\
&+& \left. 
3\,\frac{\partial^4 V}{\partial\ve\partial \phi^3}\mid_{\ve=0}\,\phi_1^2
+
6\,\frac{\partial^3 V}{\partial \phi^3}\mid_{\ve=0}\,\phi_1\,\phi_2
+
\frac{\partial^4 V}{\partial \phi^4}\mid_{\ve=0}\,\phi_1^3
\right] 
\lab{phi3eq}
\er
and so on.

 We can write all these equations, generically, as
\be
\partial^2 \phi_n+ \frac{\partial^2 V}{\partial \phi^2}\mid_{\ve=0}\, \phi_n= f_n 
\qquad\qquad\qquad\qquad n=1,2,3,\dots.
\lab{eqsforphin}
\ee

Note that if one expands $V$ in powers of $\ve$, taking into account only its explicit dependence on $\ve$ and neglecting its implicit dependence on $\ve$ through $\phi$, one gets that 
\be
V= V\mid_{\ve=0} + \ve\, \frac{\partial V}{\partial \ve}\mid_{\ve=0}+\frac{\ve^2}{2}\,\frac{\partial^2 V}{\partial \ve^2}\mid_{\ve=0}+\ldots 
\ee
where $V\mid_{\ve=0}$ is the sine-Gordon potential \rf{sgpot}.
 Thus, one finds that
\be
\frac{\partial^2 V}{\partial \phi^2}\mid_{\ve=0}=\frac{\partial^2 \(V\mid_{\ve=0}\)}{\partial \phi^2}\mid_{\ve=0}= \cos\(4\, \phi_0^{(-)}\).
\lab{d2vphi2e0}
\ee
Indeed, one can check explicitly that the potentials \rf{v1def},   \rf{v2def} and  \rf{v3def} mentioned in the last section  do satisfy \rf{d2vphi2e0}. In fact this is true in all models which involve deformations as discussed in \cite{new} as for all of them 
$\frac{\partial^2 \(V\mid_{\ve=0}\)}{\partial \phi^2}\mid_{\ve=0}= \cos\(4\, \phi_0\)$. Thus the approach of \cite{new} guarantees
\rf{d2vphi2e0}.

We now split \rf{eqsforphin} into the even and odd parts under the parity $P$, as
\be
\partial^2 \phi_n^{(\pm)}+ \cos\(4\, \phi_0^{(-)}\)\, \phi_n^{(\pm)}= f_n^{(\pm)} 
\qquad\qquad\qquad {\rm with}\qquad \quad \star^{(\pm)}\equiv \frac{1}{2}\(1\pm P\) \star
\lab{eqsforphinsplit}
\ee

Then using \rf{v1def},   \rf{v2def} and  \rf{v3def} one can verify  that
\br
f_1^{(1)}&\equiv&-\frac{\partial^2 V_1}{\partial \ve\partial\phi}\mid_{\ve=0}
\\
&=&
\frac{1}{16} \left[\left(\pi ^2-12 (\phi_0^{(-)})^2\right) \sin (4 \phi_0^{(-)} )+4 \left(4 (\phi_0^{(-)})^2-\pi
   ^2+3\right) \phi_0^{(-)}  \cos (4 \phi_0^{(-)} )-12 \phi_0^{(-)} \right],
   \nonumber\\
f_1^{(2)}&\equiv&-\frac{\partial^2 V_2}{\partial \ve\partial\phi}\mid_{\ve=0}
\\
&=&\frac{1}{8} \left[\left(-8 (\phi_0^{(-)})^2+4 \pi  \phi_0^{(-)} -2\right) \cos (4 \phi_0^{(-)} )-(\pi -4 \phi_0^{(-)} )
   \sin (4 \phi_0^{(-)} )+2\right],
    \nonumber\\
f_1^{(3)}&\equiv&-\frac{\partial^2 V_3}{\partial \ve\partial\phi}\mid_{\ve=0}
\\
&=& \frac{1}{32} \left[\left(16 \gamma \phi_0^{(-)} 
-  4 \pi  \gamma-12 (\phi_0^{(-)})^2+\pi ^2\right) \sin (4 \phi_0^{(-)} )
+8 \gamma-12 \phi_0^{(-)}
\right. \nonumber\\
&-& \left.  4
   \left(\gamma \left(8 (\phi_0^{(-)})^2-4 \pi  \phi_0^{(-)} +2\right)+\phi_0^{(-)}  \left(-4 (\phi_0^{(-)})^2+\pi
   ^2-3\right)\right) \cos (4 \phi_0^{(-)} ) \right].
    \nonumber
\er

Then we note that $\frac{1}{2}\(1- P\) f_1^{(j)} \neq 0$, for $j=1,2,3$, and that 
\br
\frac{1}{2}\(1+ P\) f_1^{(1)} &=& 0,
\nonumber\\
\frac{1}{2}\(1+ P\) f_1^{(2)} &\neq& 0,
\nonumber\\
\frac{1}{2}\(1+ P\) f_1^{(3)} &=&\biggl\{ \begin{array}{c}
 0 \qquad \quad {\rm for}\qquad \gamma=0,\\
\neq 0 \qquad \quad {\rm for}\qquad \gamma\neq 0.
\end{array}
\nonumber
\er

Thus we see that for the theories described by the potential $V_1$, given in \rf{v1def}, and the potential $V_3$, given in \rf{v3def}, for $\gamma =0$, the even part (under the parity) of the first order field $\phi_1$ satisfies homogeneous equations, and the odd part of the first order field $\phi_1$ satisfies non-homogeneous equations, {\it i.e.}
\br
\partial^2 \phi_1^{(+)}+ \frac{\partial^2 V}{\partial \phi^2}\mid_{\ve=0}\, \phi_1^{(+)}&=& 0,
\\
\partial^2 \phi_1^{(-)}+ \frac{\partial^2 V}{\partial \phi^2}\mid_{\ve=0}\, \phi_1^{(-)}&\neq & 0 \qquad \quad {\rm for} \qquad V_1 \quad {\rm and} \quad V_3\mid_{\gamma=0}.
\nonumber
\er

Thus, for such cases, one can always choose $\phi_1^{(+)}=0$. Indeed, if $\phi_1$ is a solution, so is $\phi_1=\phi_1-\phi_1^{(+)}=\phi_1^{(-)}$. 

One can also check the if the choice $\phi_1^{(+)}=0$ is made, then it turns out the for theories governed by $V_1$ and $V_3\mid_{\gamma=0}$ the second order field $\phi_2=0$ has similar properties, {\it i.e.} its even part under the parity satisfies homogenous equations, and its odd part non-homogeneous equations. So, one can similarly take $\phi_2^{(+)}=0$. 

Continuing this process one reaches the conclusion that it is always possible to have two-soliton-like and breather-like solutions for the theories with potentials $V_1$ and $V_3\mid_{\gamma=0}$, which are odd under the parity $P$. In addition, one can check that when evaluated on such odd solutions the potentials $V_1$ and $V_3\mid_{\gamma=0}$ are even under the parity $P$. Therefore, such theories satisfy the requirements \rf{require1} and \rf{require2} and so they possess an infinite number of charges satisfying the property \rf{mirrorcharge}, {\it i.e.} such theories are, using our definition,  {\em quasi-integrable}. 

The same argument does not apply to the theories governed by the potentials $V_2$ and $V_3\mid_{\gamma\neq 0}$, and so they cannot be considered quasi-integrable. Of course, it still may be the case, that they satisfy \rf{quasi} for other reasons.

\section{Numerical simulations}
\label{sec:numerical}
\setcounter{equation}{0}

To perform the numerical simulations to check our hypothesis  we still have to construct two-soliton configurations in our  models.

In our numerical simulations we want to analyze the quasi-conservation of the first non-trivial charge beyond the energy itself, {\it i.e.} the charge $Q^{(3)}$ defined in \rf{chargeanomalydef}. We  do this by computing the corresponding anomaly $\alpha^{(3)}$ also introduced  in \rf{chargeanomalydef}.  One can check that the factor $\gamma^{(3)}$ entering in the definition of $\alpha^{(3)}$ is given by (see \cite{us}) 
\be
\gamma^{(3)}= i\,\omega\, \partial_{-}^2 \phi 
\ee
with $\partial_{-}=\partial_t-\partial_x$. Thus, using \rf{chargeanomalydef} and \rf{xdef}, one finds that
\be
\alpha^{(3)}=-8 \, \int_{-\infty}^{\infty} dx\, \partial_{-} \phi\,\partial_{-}^2 \phi\,\left[\frac{d^2\,V}{d\,\phi^2}+16\, V-1\right],
\ee
where we have chosen $\omega =4$ and $m=1$, which are the values that make $X$ vanish when $V$ is the Sine-Gordon potential (see comments around \rf{sgpot}). We also want to compute the so-called integrated anomaly given by (see \rf{timedercharges}) 
\be
\beta^{(3)}= -\frac{1}{2}\,\int_{t_0}^t dt^{\prime} \, \alpha^{(3)} = Q^{(3)}\(t\)- Q^{(3)}\(t_0\),
\ee
where $t_0$ is the initial time of the simulation, usually taken to be zero. 

It is easy to check for the potentials of the form \rf{one.three} one has that 
\be
\frac{d^2\,V}{d\,\phi^2}+16\, V-1=2\left[\frac{\partial}{\partial\phi}\left(\,G\frac{\partial G}{\partial \phi}\right)\,+\,2G^2-2\right]V_0\,+\,3 \frac{dV_0}{d\psi}\frac{\partial G}{\partial \phi},
\label{two.aa}
\ee
where $V_0=\sin(\psi(\phi))$ and $G=\frac{\partial \phi}{\partial \psi}$.

Note that in the case of the Sine-Gordon potential, which corresponds to $\psi\(\phi\)=\phi$ and so $G=1$,  one sees that this quantity indeed vanishes.


\subsection{General considerations}

Our numerical simulations were performed using the 4th order Runge-Kutta method of simulating time evolution.
We experimented with various grid sizes and numbers of points and most simulations were performed on lattices of 10001 lattice
points with lattice spacing of 0.01  (so they covered the region of (-50.0,\,50.0). Time step $dt$ was 0.0001. 
The solitons (for the soliton-soliton scattering) were placed at $\pm 35.00$ and stretched $\pm 5.00$ from their positions hence at the
edges of the grid the fields resembled the vacuum configurations which were modified only by waves that were emitted during the scattering. The same was true for the breather-like structures - except that this time the field configurations were different from their vacuum 
values only for a very small range of $x$ ({\it i.e.} for $\vert x\vert <20$).

At the edges of the grid ({\it i.e.} for $49.50<\vert x\vert <50.00$) we absorbed the waves reaching this region (by decreasing 
the time change of the magnitude of the field there). 

In consequences, the total energy was not conserved but the only energy which was absorbed was the energy of radiation waves.
Hence the total energy was  effectively the energy of the field configuration which we wanted to study.

As we will discuss later in this section the energy was very well conserved in all soliton - soliton scatterings and much less 
so for the evolution of our breather-like structures.

\subsection{Two soliton configurations}

For the first model we note that we can put one soliton $\psi$ to lie between ($-\frac{\pi}{2},\,0)$ and the second one to lie between $(0,\,\frac{\pi}{2})$.
Thus our initial condition is given by ($f$ stands for the initial field configuration) 
\be
f\,=\,\frac{g}{1+\varepsilon\pi^2-\varepsilon g^2},
\label{two.eleven}
\ee
where 
\be
g\,=\,-\,{\rm ArcTan}\left[\exp\left(-\frac{-(x+a_1)+vt}{\sqrt{1-v^2}}\right)\right]\quad \hbox{for}\quad x<0
\label{two.twelve}
\ee
and
\be
g\,=\,\,{\rm ArcTan}\left[\exp\left(-\frac{x-a_1+vt}{\sqrt{1-v^2}}\right)\right]\quad \hbox{for}\quad x>0,
\label{two.thirteen}
\ee
and where $a_1>0$.

In this case one soliton is placed at $-a_1$ and the other one at $a_1$. For $x<0$ $f(x,0)$ varies from $-\frac{\pi}{2}$ (for $x\sim-\infty$) to $f\sim0$ for $x=0$. For $x>0$ $f(x,0)$ varies from 0 (for $x\sim0$) to $f=\frac{\pi}{2}$ for $x\sim \infty$.

In the second case we take

\be 
f\,=\,\frac{1+\varepsilon\frac{ \pi}{2}g}{1+\varepsilon g},
\label{two.fourteen}
\ee
where 
\be
g\,=\,{\rm ArcTan}\left[\exp\left(\frac{(x+a_1)-vt}{\sqrt{1-v^2}}\right)\right]\quad \hbox{for}\quad x<0
\label{two.fifteen}
\ee
and
\be
g\,=\,{\rm ArcTan}\left[\exp\left(\frac{x-a_1+vt}{\sqrt{1-v^2}}\right)\right]\,+\,\frac{\pi}{2}\quad \hbox{for}\quad x>0,
\label{two.sixteen}
\ee
and where again $a_1>0$.

The two solitons are again placed at $\pm a_1$ but this time for $x<0$ $f(x,0)$ varies from $0$ (for $x\sim-\infty$) to $f\sim \frac{\pi}{2}$ for $x=0$. For $x>0$ $f(x,0)$ varies from $\frac{\pi}{2}$ for $x\sim0$ to $f\sim\frac{\pi(1+\varepsilon\frac{ \pi}{2})}{1+\varepsilon\pi}$ for $x\sim \infty$.

\begin{figure}
\begin{center}
\includegraphics[width=0.3\textwidth]{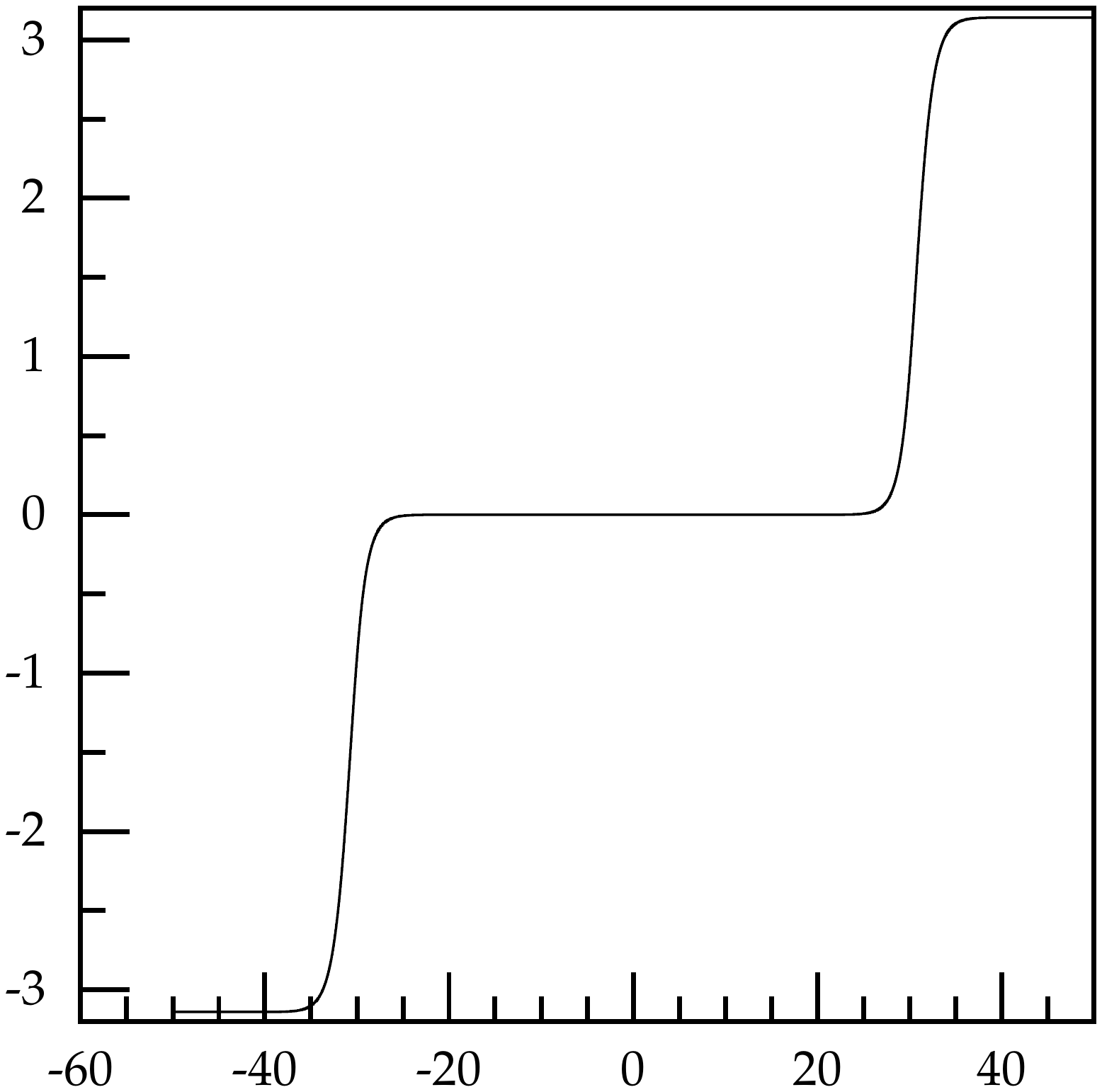} 
\includegraphics[width=0.3 \textwidth]{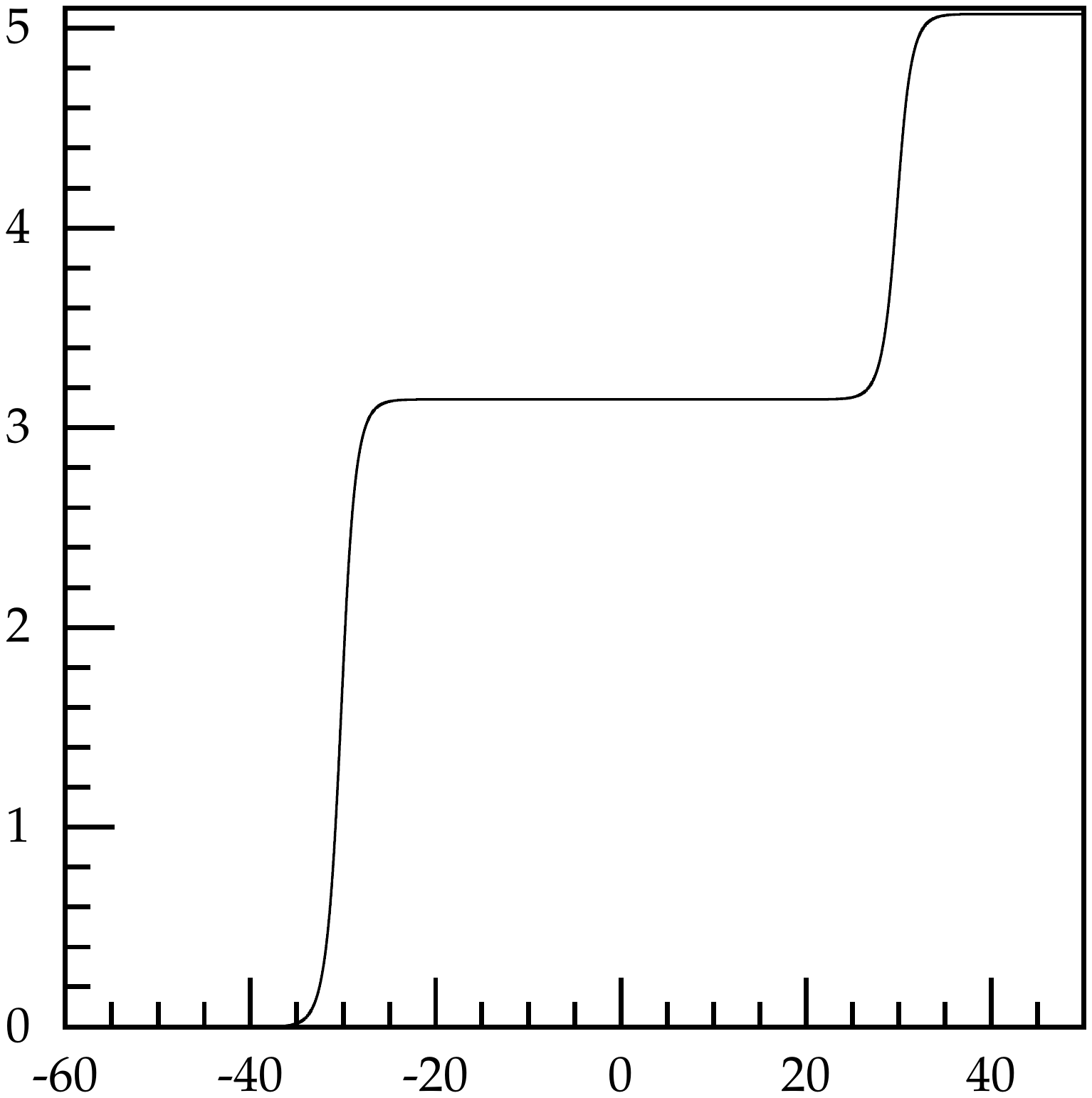}  
\end{center}
\caption[AS]{\small The initial fields of two solitons in the first two classes of models ($\varepsilon=0.1$, $v=0.05$). In fact the plotted values are $2\phi(0,x)$}.
\label{fig1}
\end{figure}

In fig \ref{fig1} we plot the fields at $t=0$ used in our simulations. The values of $\varepsilon$, $a_1$ and $v$ are $\varepsilon=0.1$ $a_1=30.0$ and $v=0.05$.

We note that the first model possesses the required symmetry while the second one does not.

Next we have performed simulations of the scatterings and the anomalies in each model 
for many values of all parameters.

\begin{figure}
\begin{center}
\includegraphics[width=0.3\textwidth]{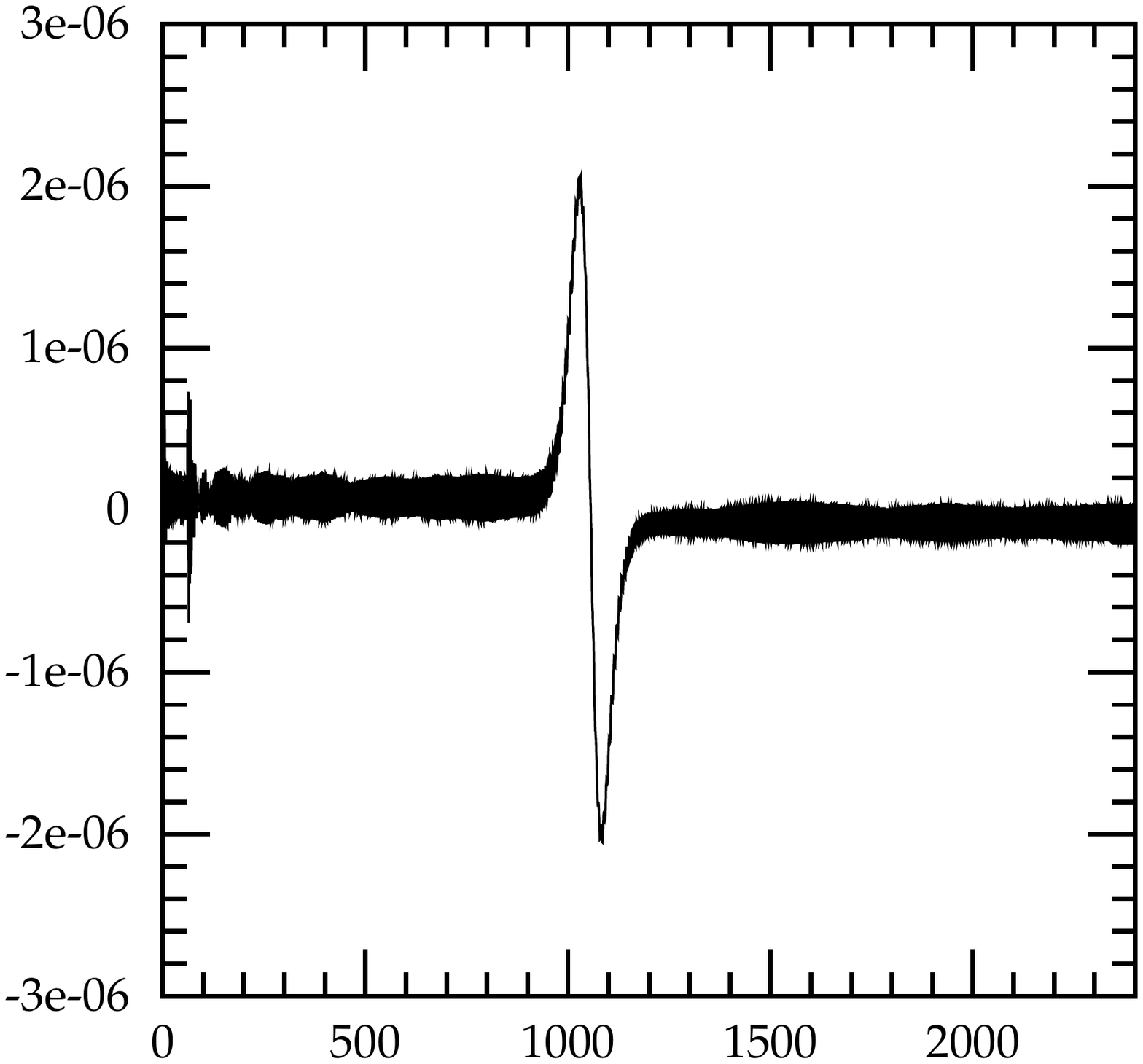} 
\includegraphics[width=0.3 \textwidth]{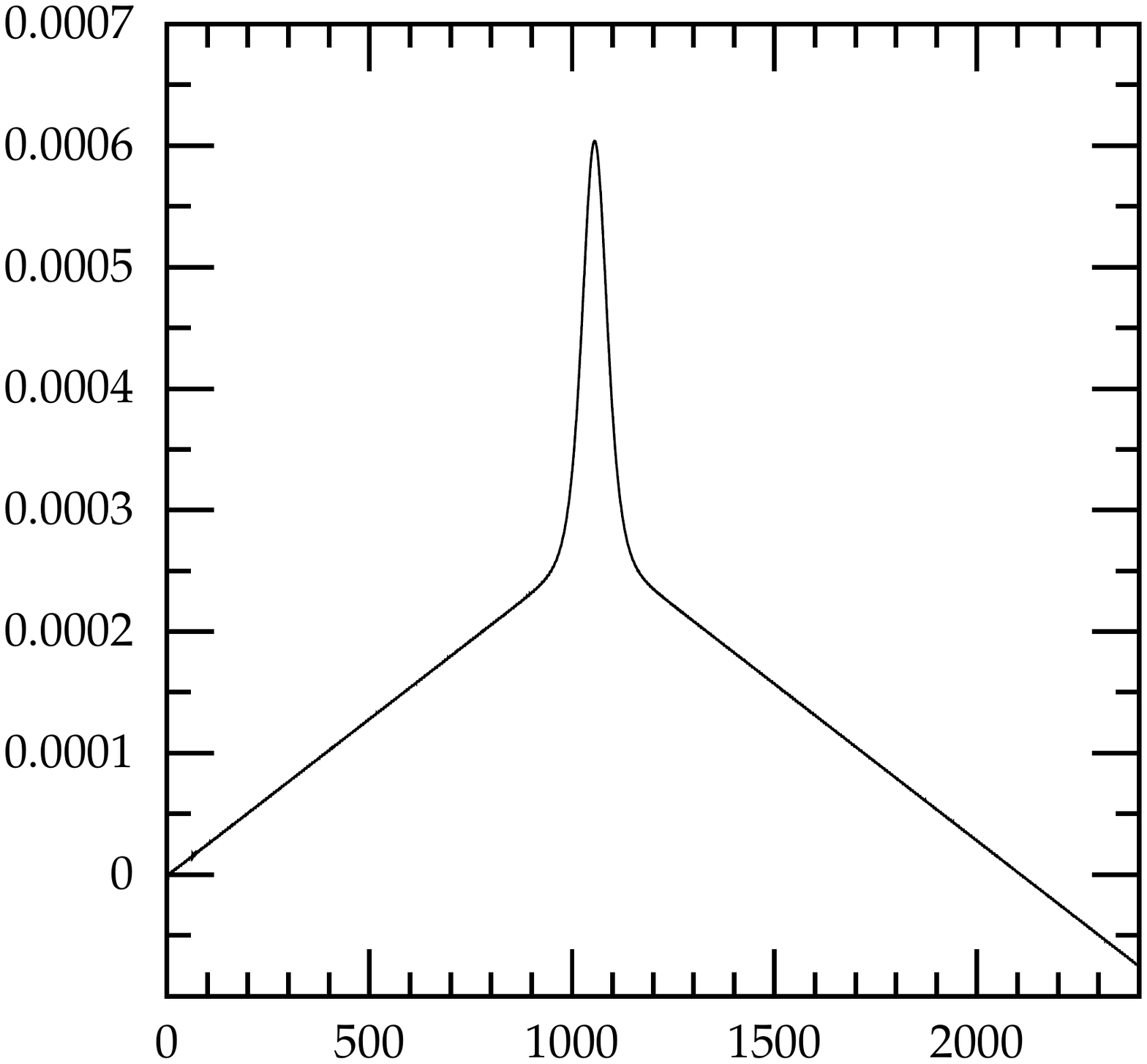}  
\end{center}
\caption[AS]{\small Anomaly (a) and time integrated anomaly (b) seen in the first class of models for $\varepsilon=0.01$ and $v=0.025$.}
\label{fig2}
\end{figure}

\begin{figure}
\begin{center}
\includegraphics[width=0.3\textwidth]{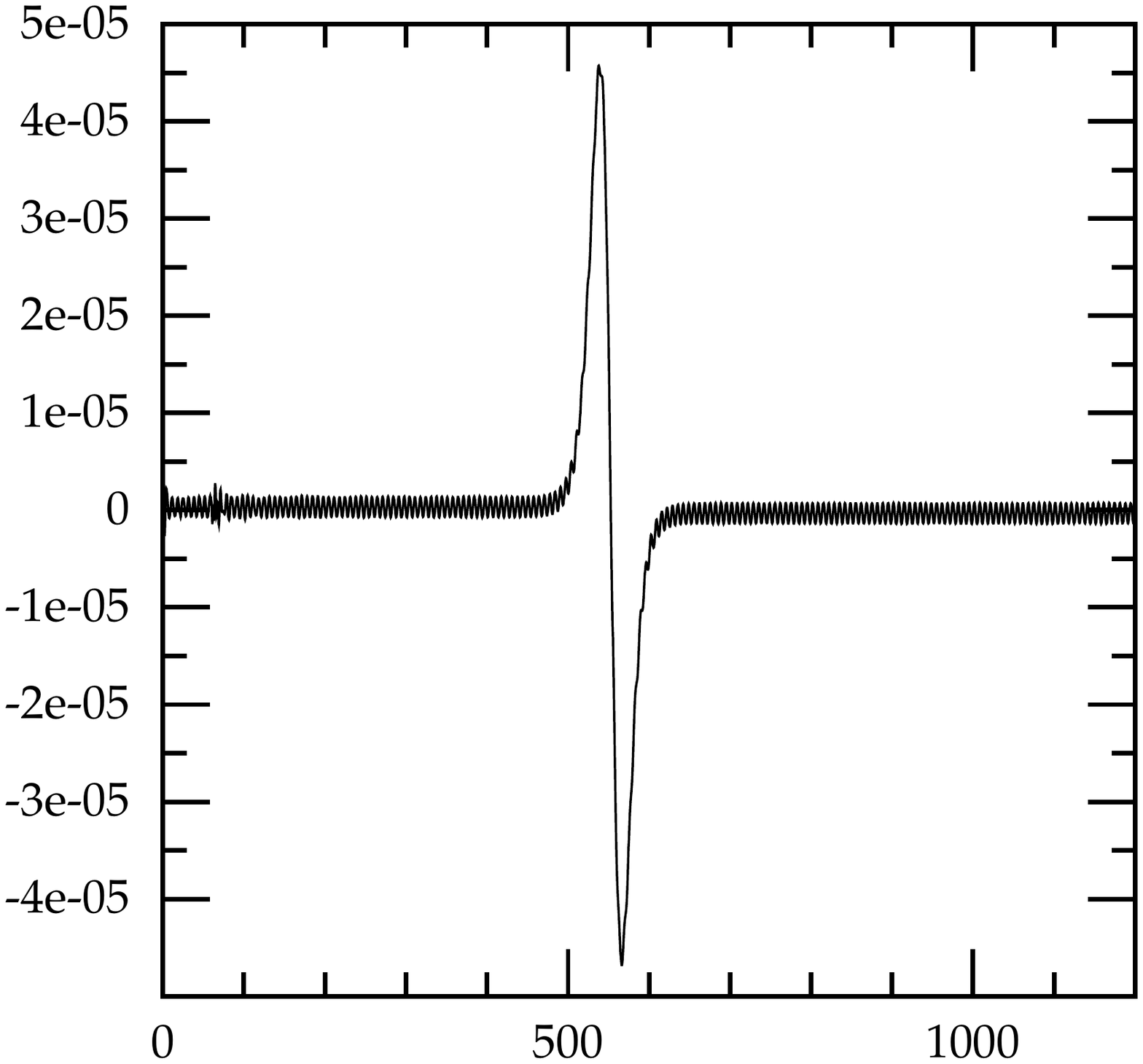} 
\includegraphics[width=0.3 \textwidth]{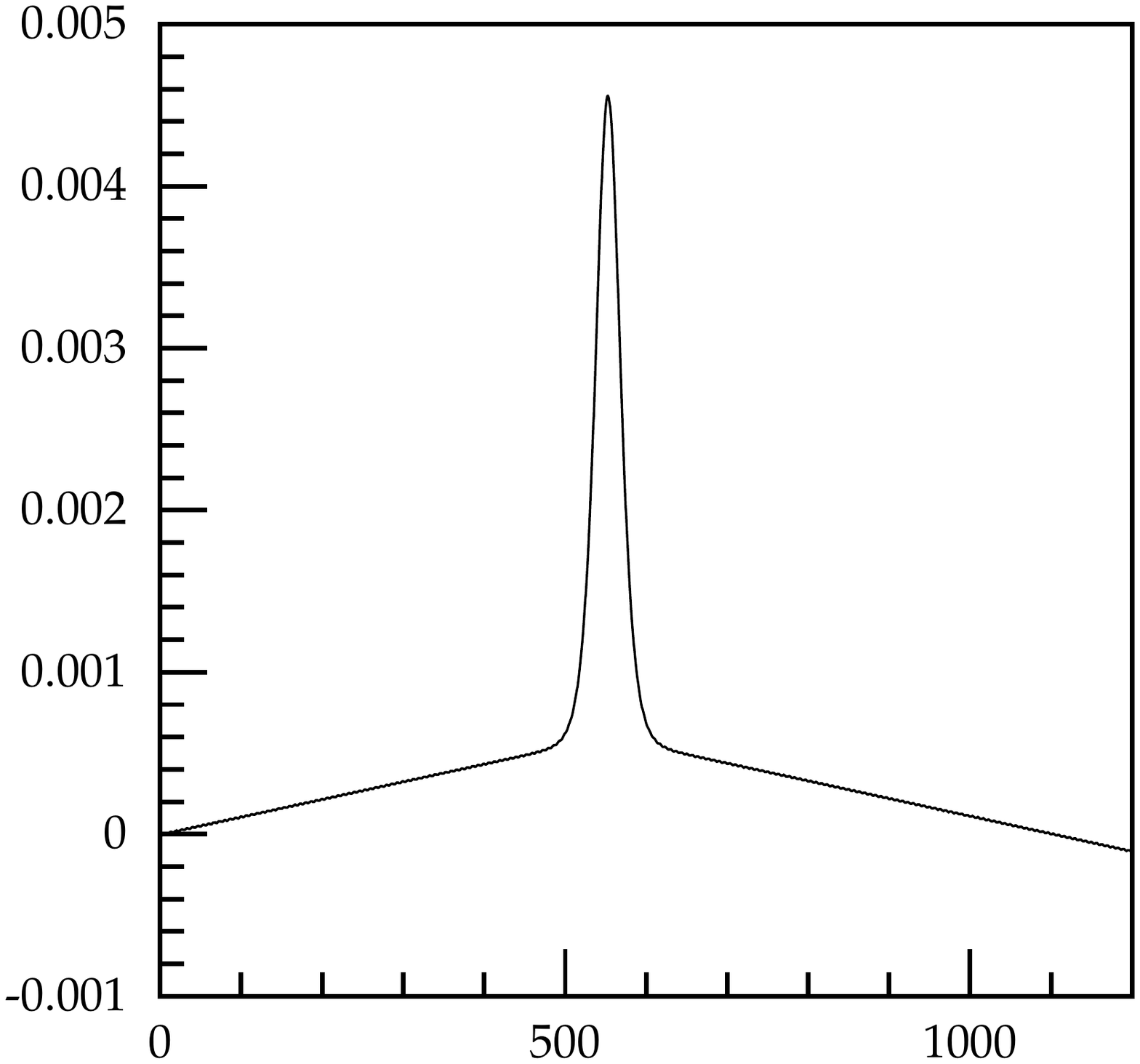}  
\end{center}
 \caption[AS]{\small As fig. \ref{fig2} but for $\varepsilon=0.1$ and $v=0.05$.}
\label{fig3}
\end{figure}
 
\begin{figure}
\begin{center}
\includegraphics[width=0.3\textwidth]{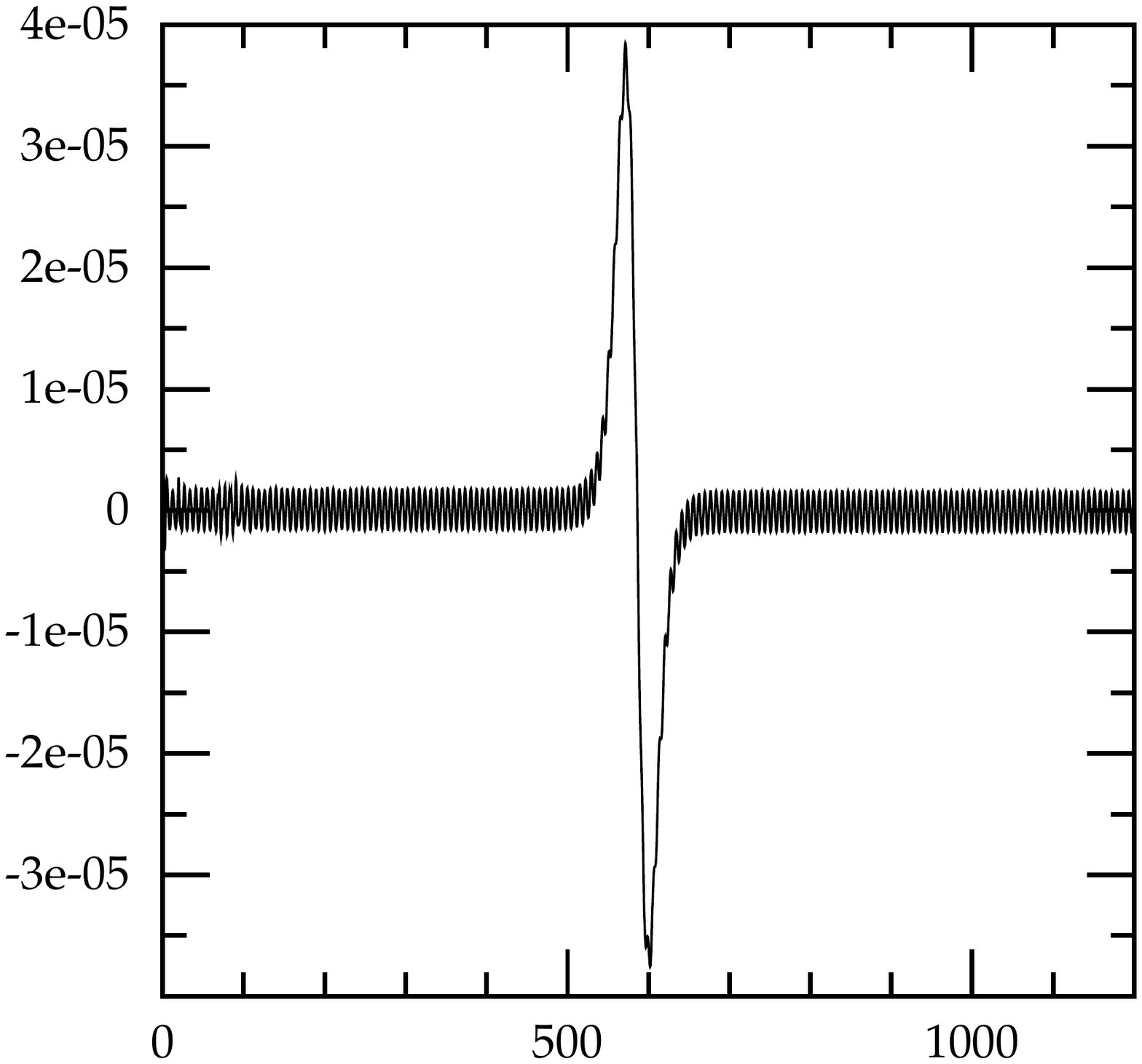} 
\includegraphics[width=0.3 \textwidth]{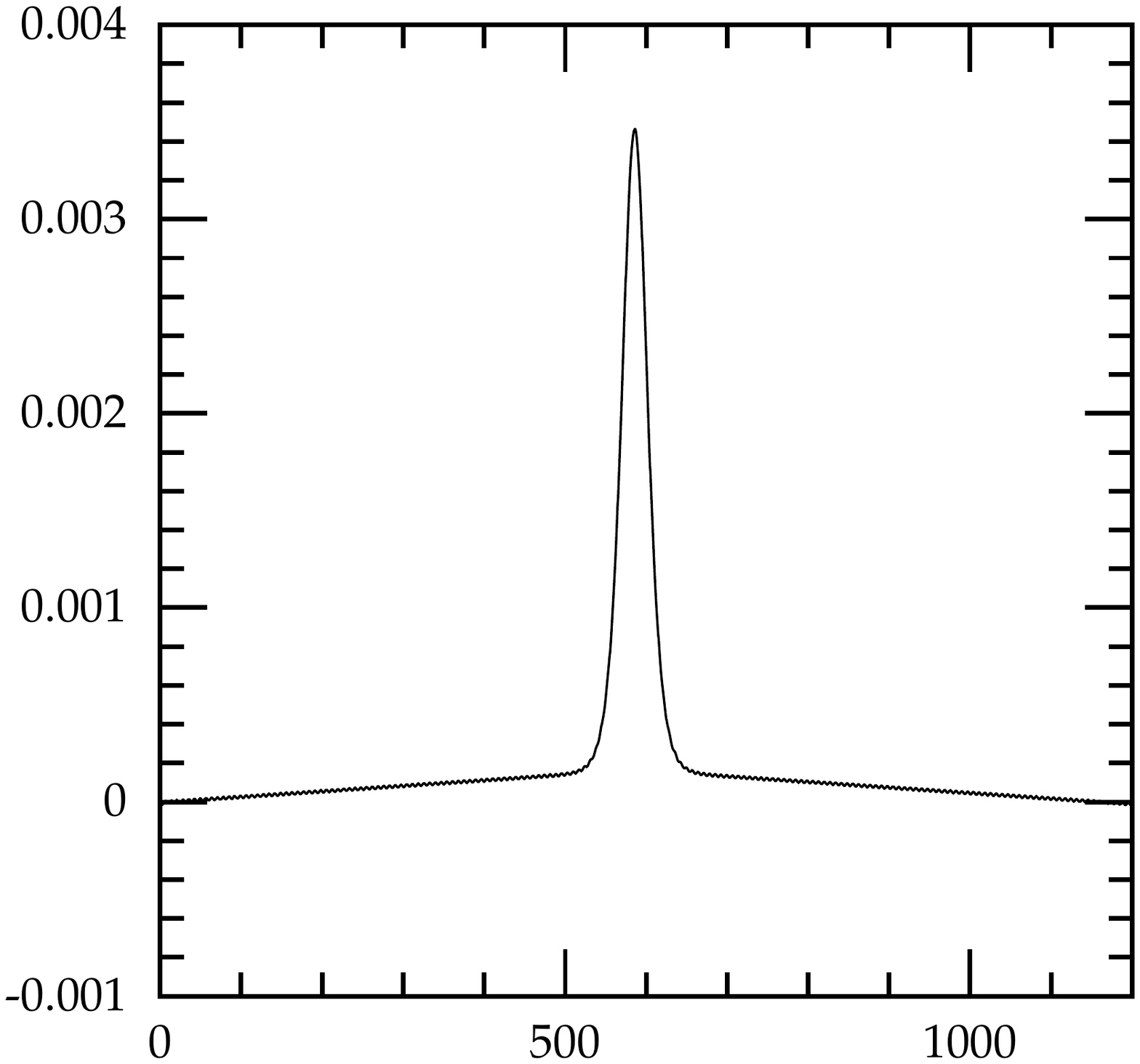}  
\end{center}
 \caption[AS]{\small As fig. \ref{fig2} but for $\varepsilon=1.$ and $v=0.05$.} 
\label{fig4}
\end{figure}

Below we present some of our results.

First we present the results in the first model.
In figures \ref{fig2}, \ref{fig3}, \ref{fig4} we present the expressions of the anomaly (and time integrated)
for 3 values of $\varepsilon$ ($\varepsilon=0.01$ in fig. \ref{fig2}, $\varepsilon=0.1$ in fig. \ref{fig3} and $\varepsilon=1.0$ in fig. \ref{fig4})
The left-hand plots present the time dependence of anomaly; while the plots to the right of them show
the time integrated anomaly. We note that the unintegrated anomaly increases as the solitons come together, then vanishes 
when the solitons reflect and changes sign when they are going moving away from each other.
This change is responsible for the almost complete cancelation of the total anomaly and so is `generates' quasi-integrability.

\begin{figure}
\begin{center}
\includegraphics[width=0.3\textwidth]{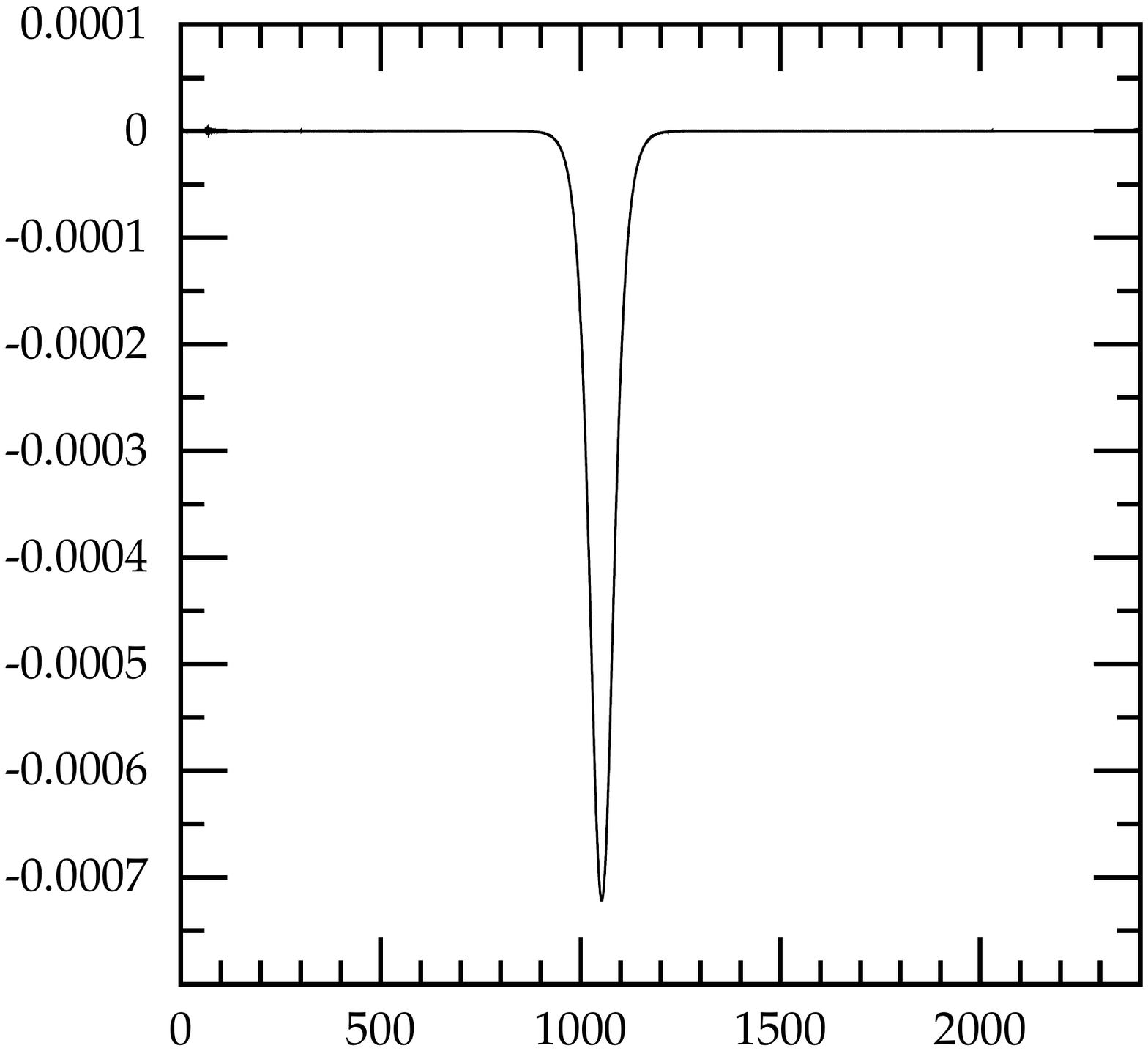} 
\includegraphics[width=0.3 \textwidth]{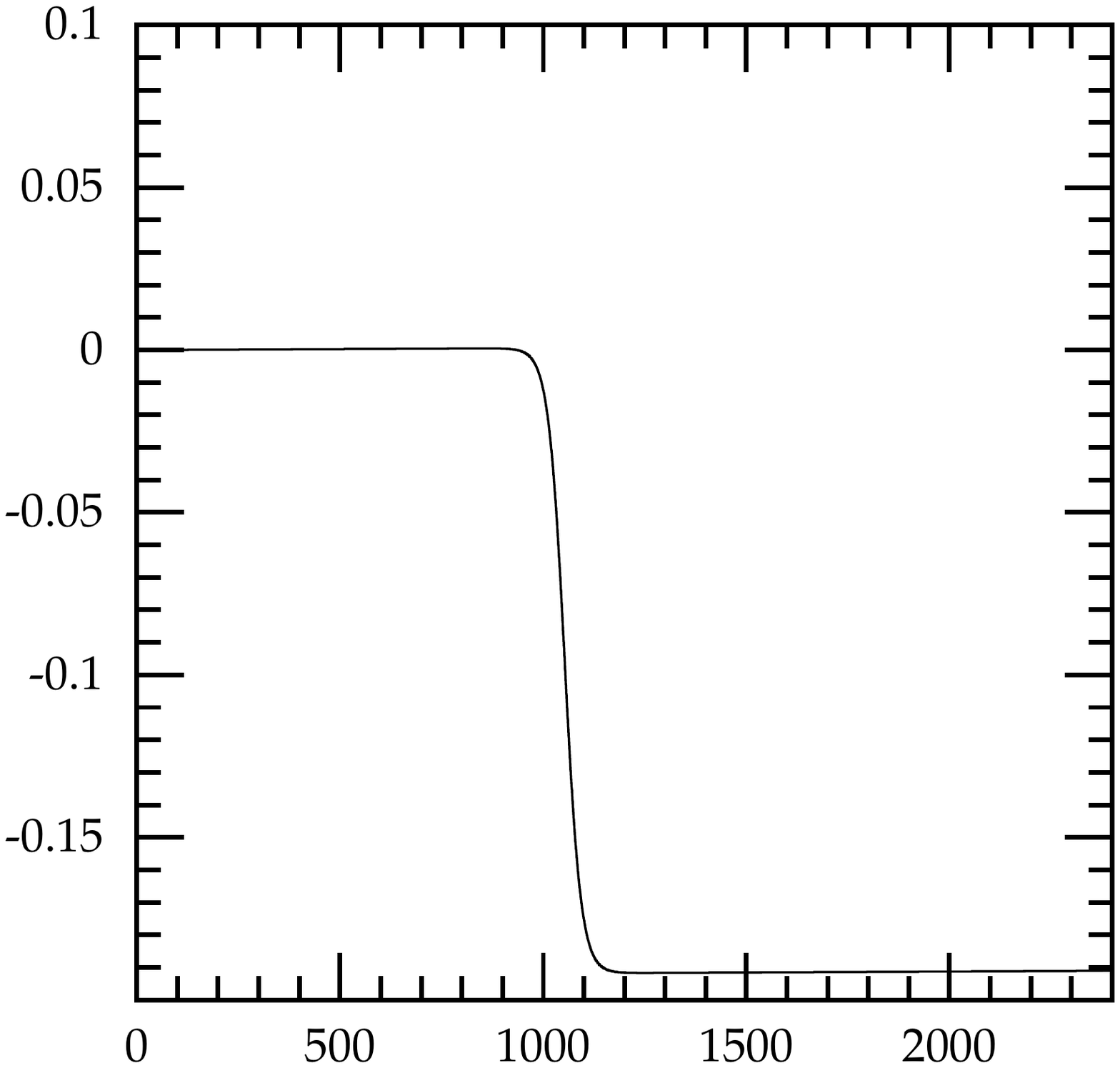}  
\end{center}
 \caption[AS]{\small Anomaly (a) and time integrated anomaly (b) seen in the second class of models for $\varepsilon=0.01$ and $v=0.025$.} 
\label{fig5}
\end{figure}

\begin{figure}
\begin{center}
\includegraphics[width=0.3\textwidth]{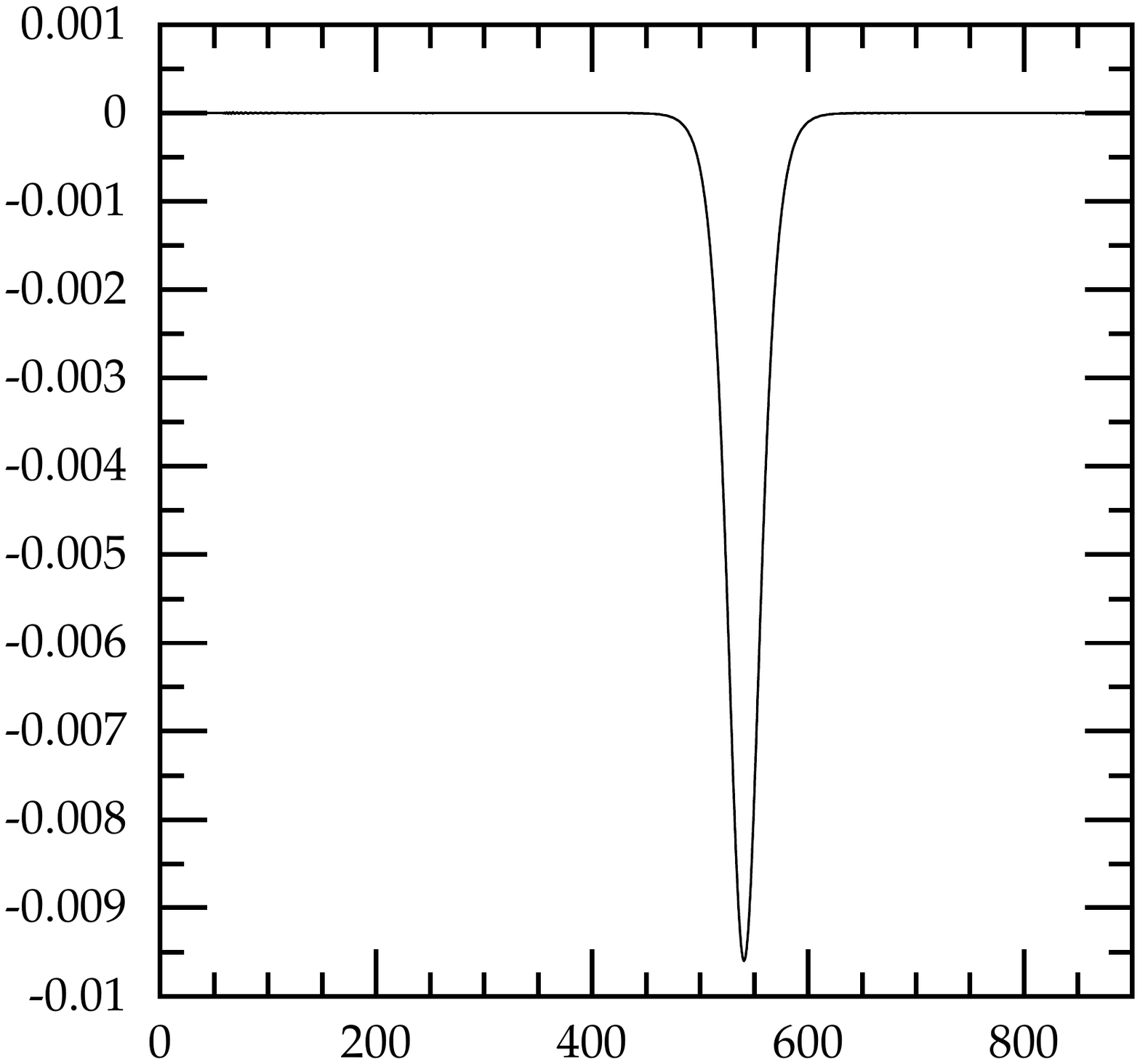} 
\includegraphics[width=0.3 \textwidth]{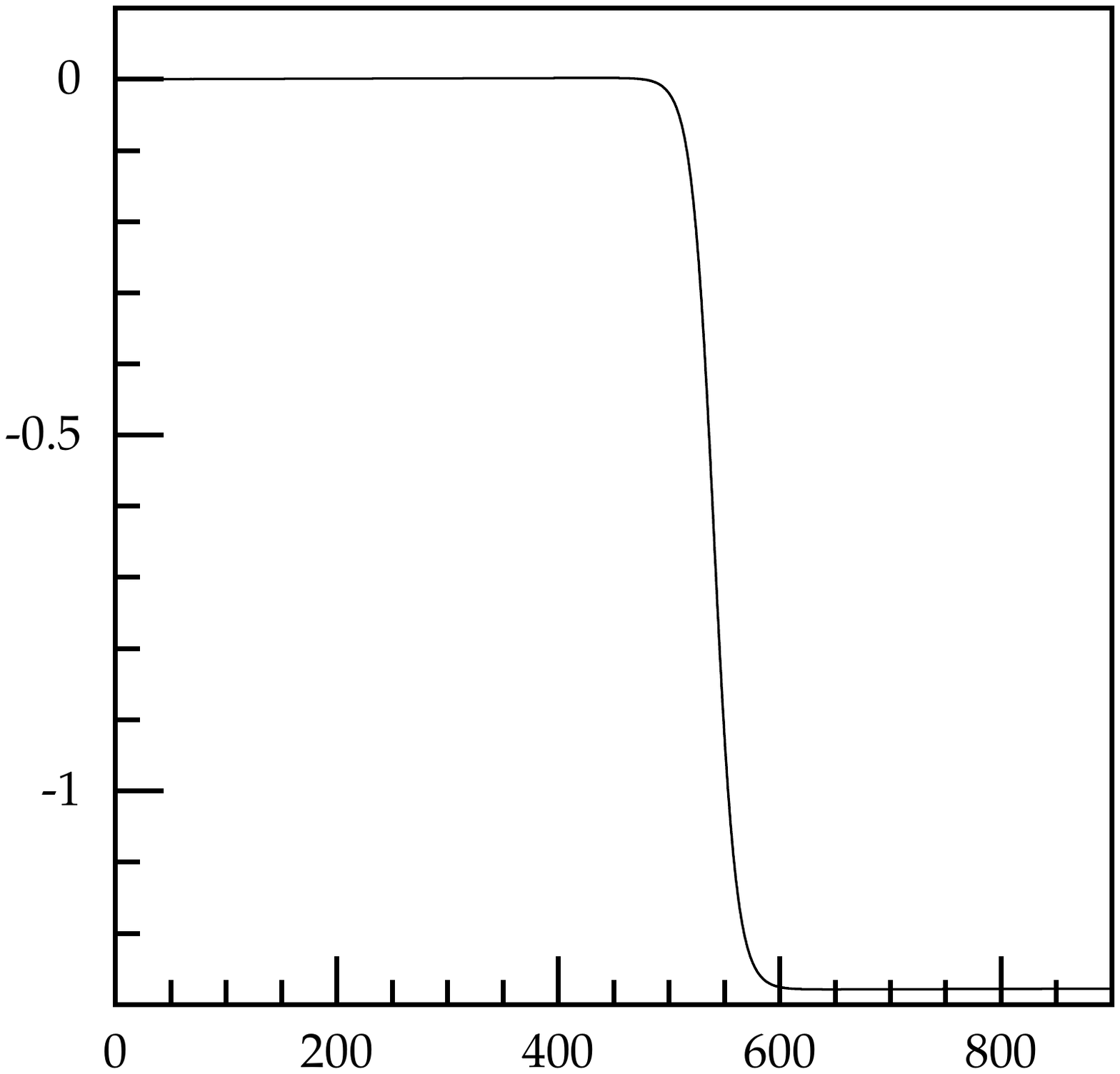}  
\end{center}
 \caption[AS]{\small As fig. \ref{fig5} but for $\varepsilon=.1$ and $v=0.05$.}
 \label{fig6}
\end{figure}

\begin{figure}
\begin{center}
\includegraphics[width=0.3\textwidth]{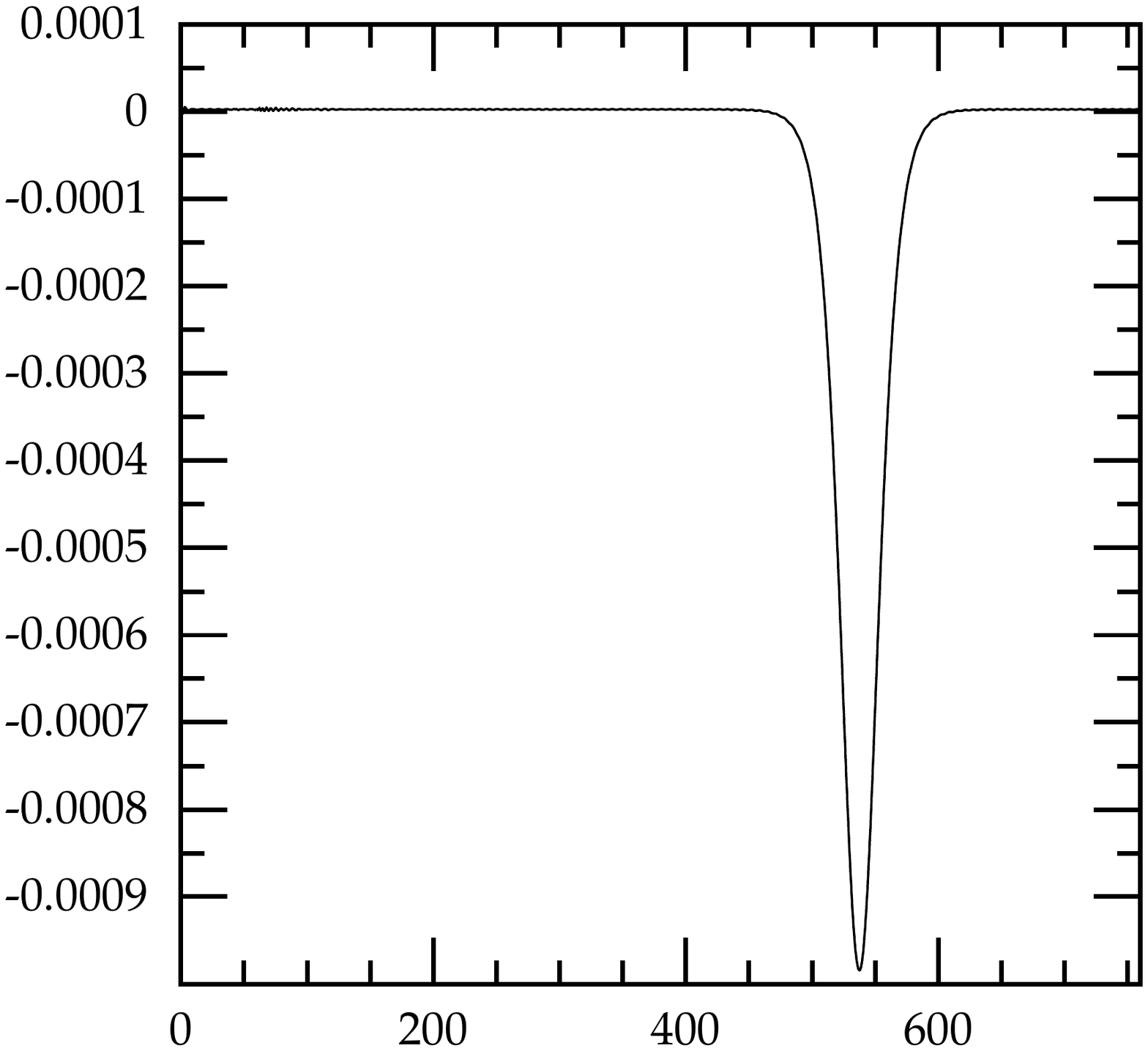} 
\includegraphics[width=0.3 \textwidth]{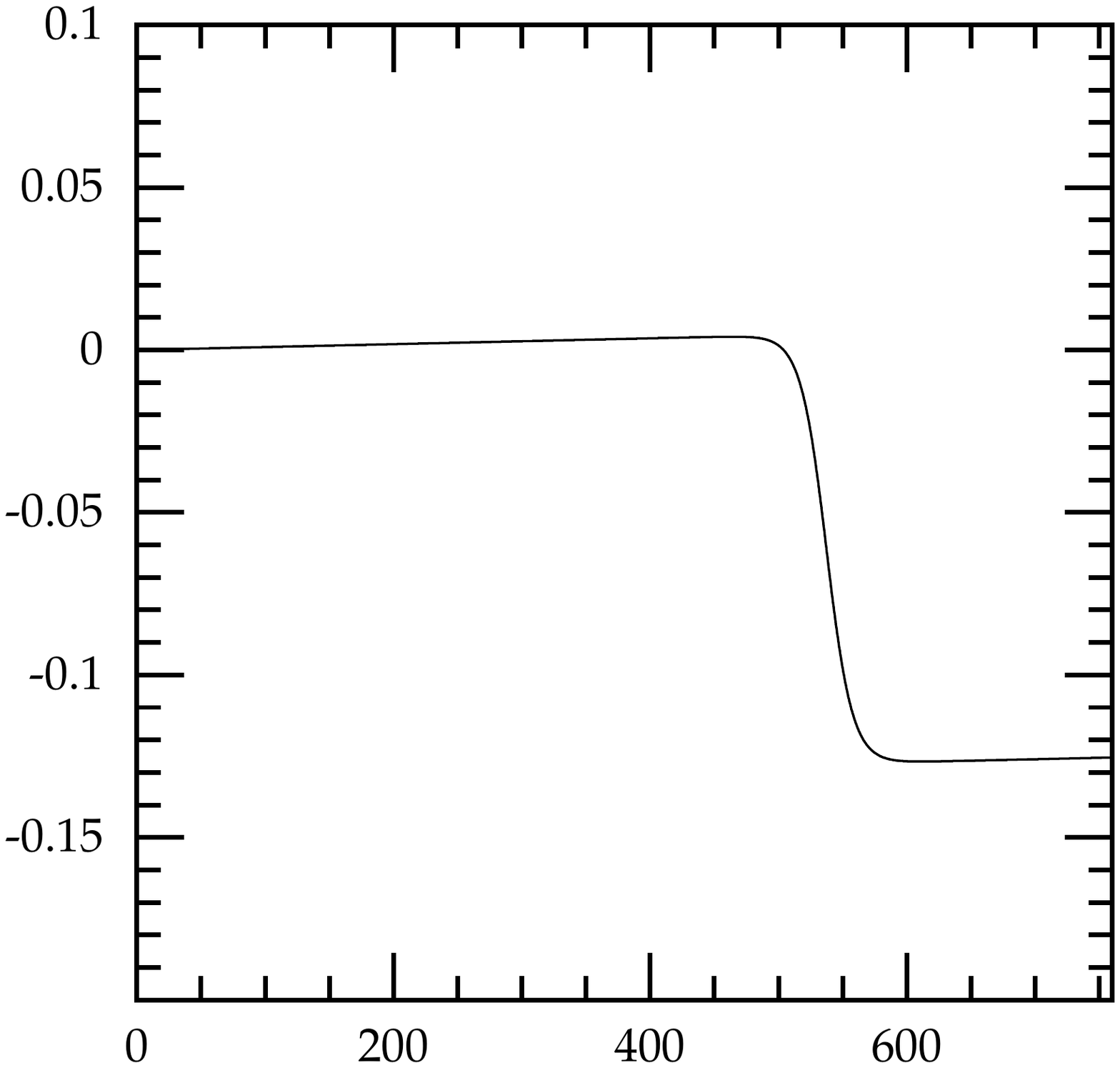}  
\end{center}
\caption[AS]{\small As fig. \ref{fig5} but for $\varepsilon=1.$ and $v=0.05$.} 
\label{fig7}
\end{figure}

Then we performed the simulation of the models of the second class. The results are presented in fig \ref{fig5}, \ref{fig6} and \ref{fig7}.
We note that this time the nonintegrated anomaly behaves very differently. Not only it stayed of only one sign but is in fact maximal when the tso solitons come to rest before reflecting. Hence, the integrated anomaly changes dramatically when the two 
solitons are very close together.

 This was the reason why  we decided to look at the possible behaviours of the unintegrated anomaly 
for a more general class of models and we have introduced the third model.

\begin{figure}
\begin{center}
\includegraphics[width=0.3\textwidth]{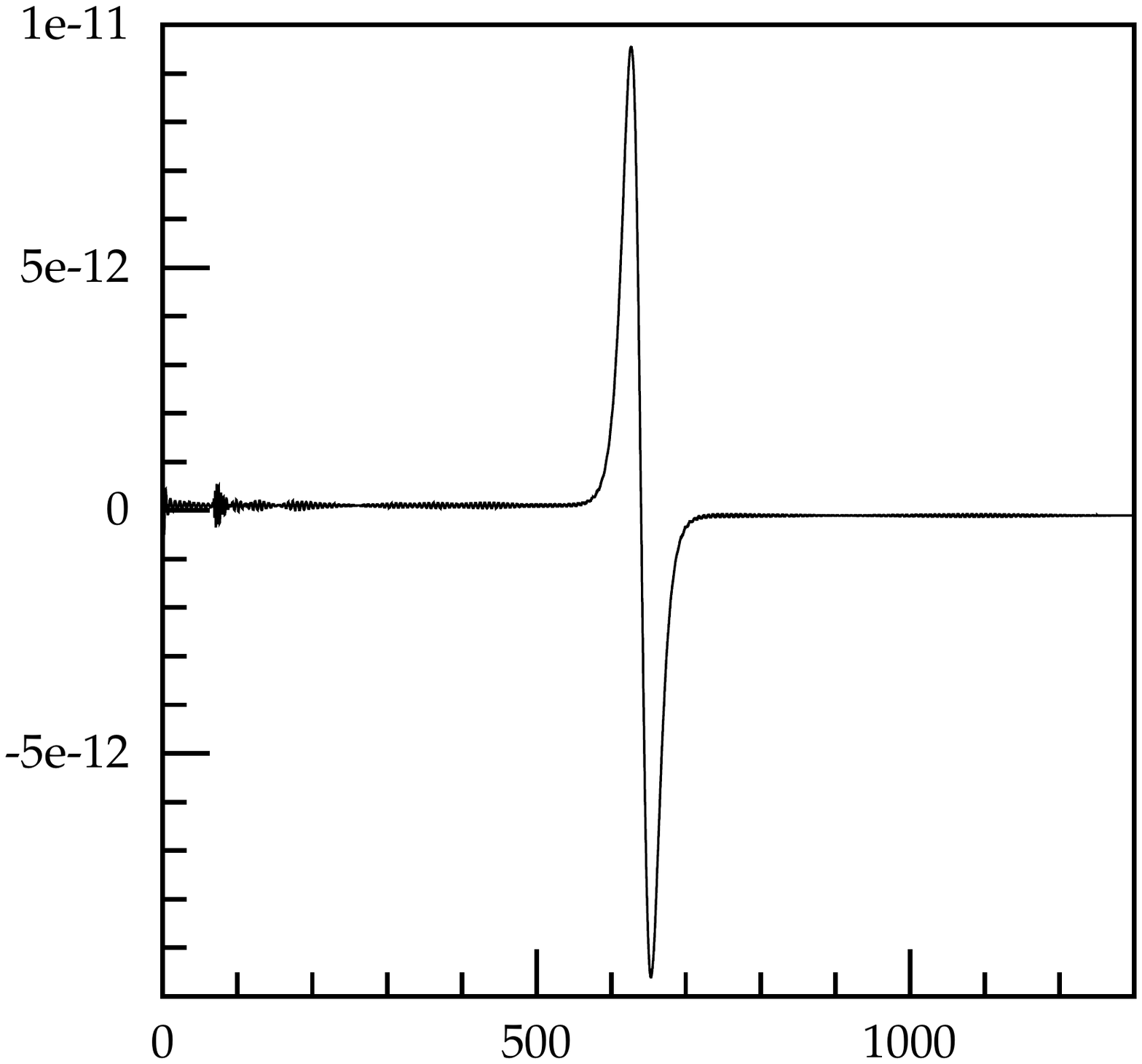} 
\includegraphics[width=0.3 \textwidth]{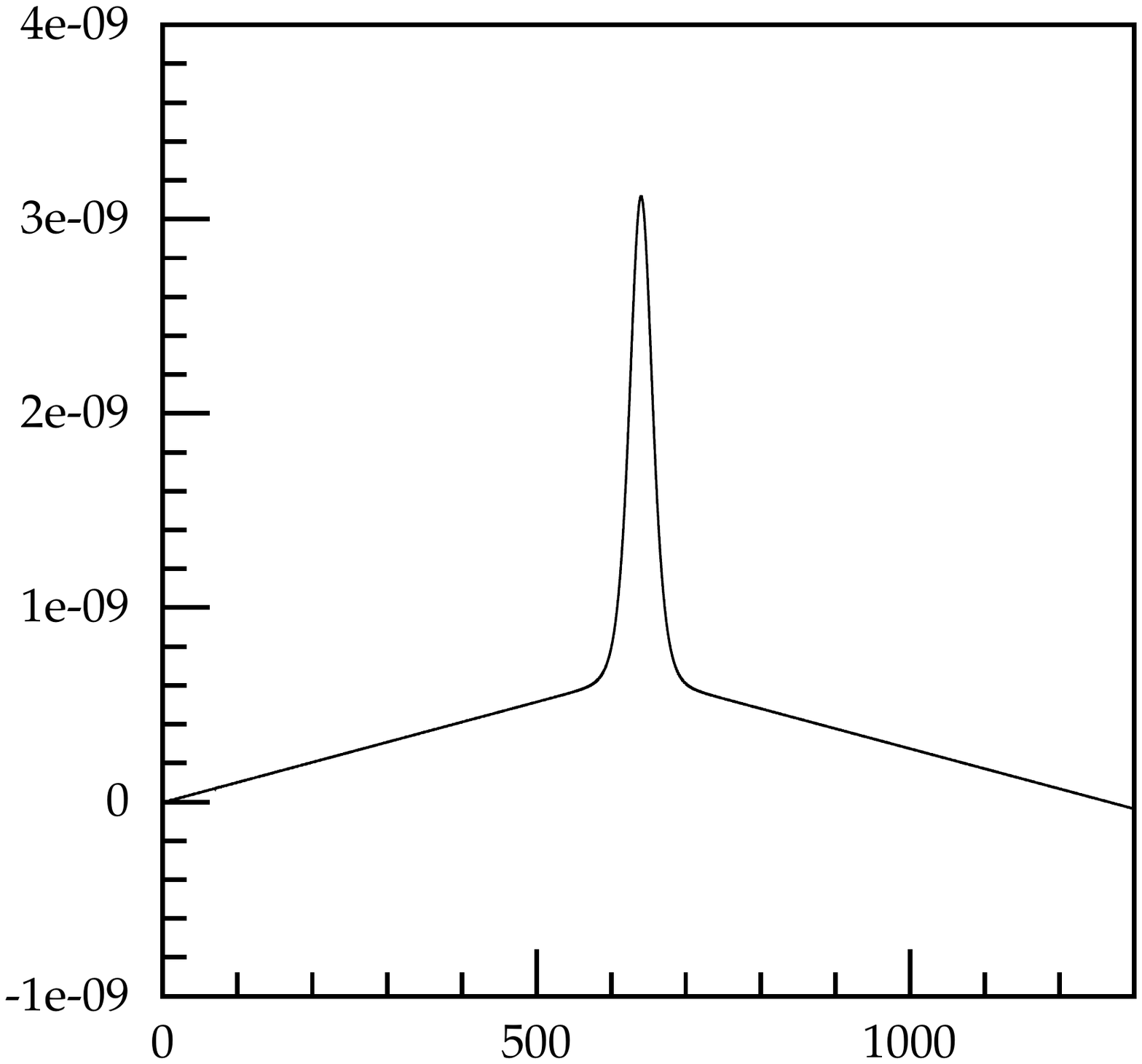}  
\end{center}
\caption[AS]{\small As fig. \ref{fig5} but for the third model. $\varepsilon=1\times 10^{-8}$, $v=0.05$ and $\gamma=1.\times 10^{-5}$.} 
\label{fig8}
\end{figure}

\begin{figure}
\begin{center}
\includegraphics[width=0.3\textwidth]{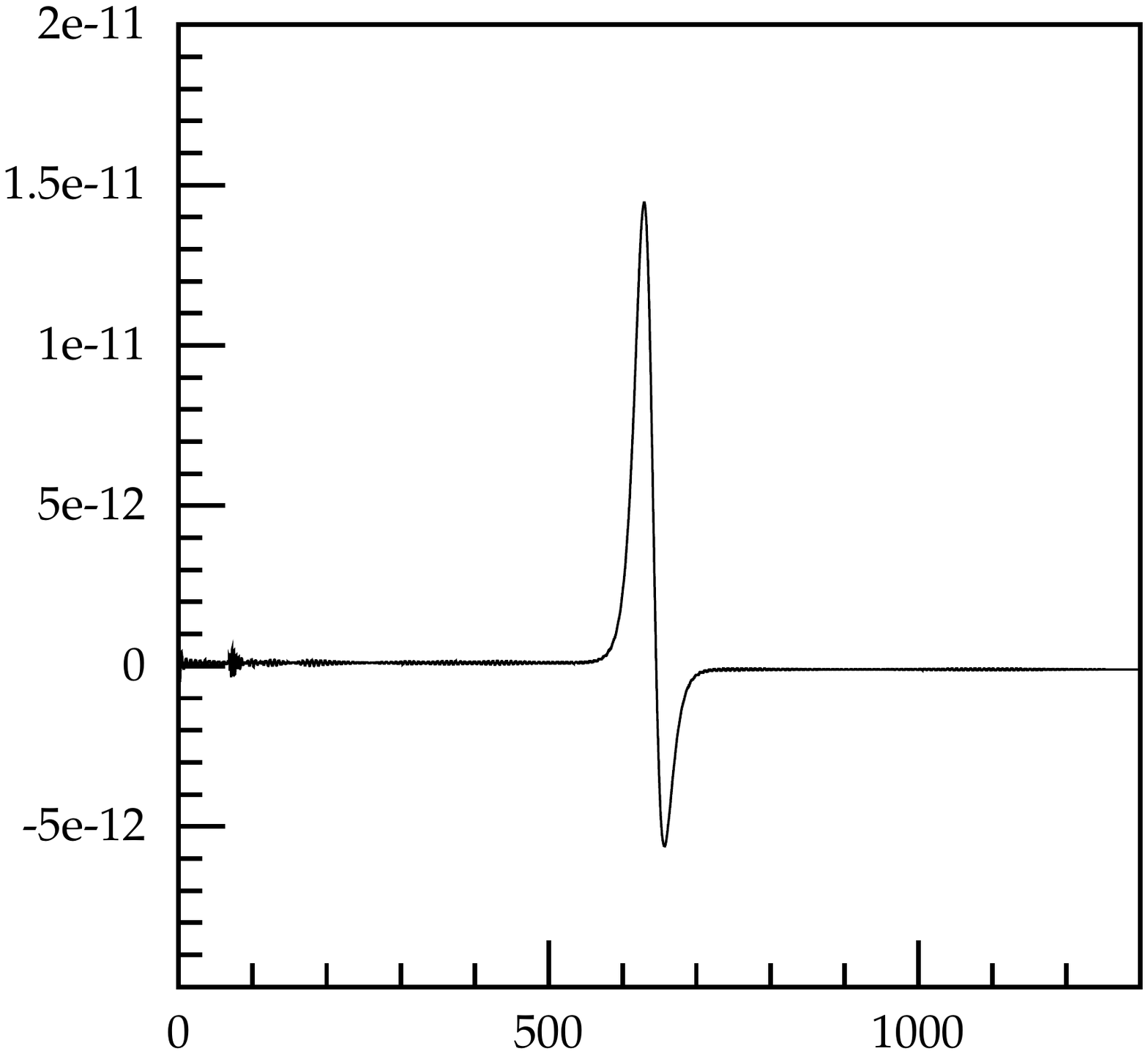} 
\includegraphics[width=0.3 \textwidth]{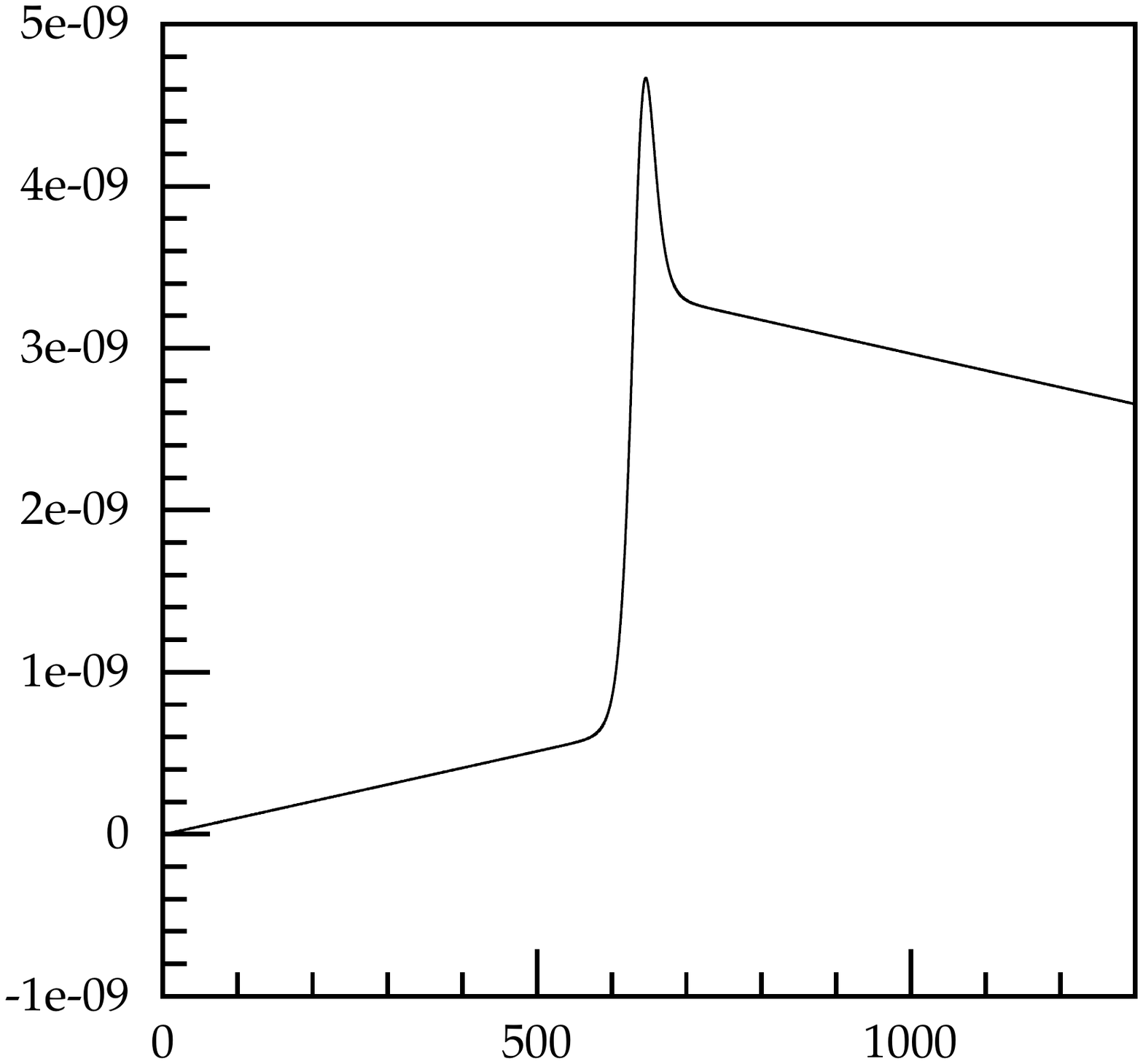}  
\end{center}
 \caption[AS]{\small As fig. \ref{fig8} but for $\gamma=-0.002$.} 
\label{fig9}
\end{figure}

\begin{figure}
\begin{center}
\includegraphics[width=0.3\textwidth]{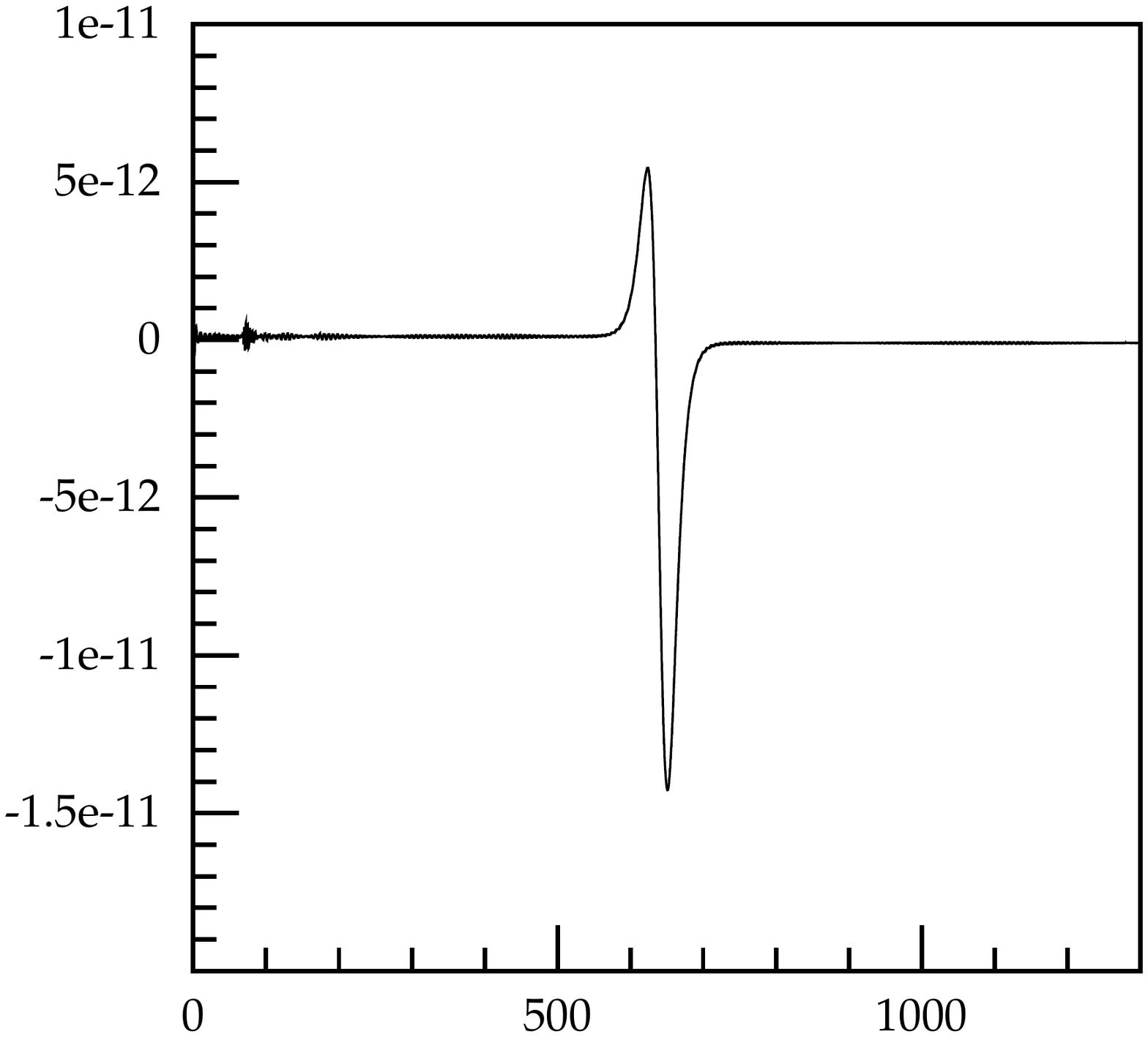} 
\includegraphics[width=0.3 \textwidth]{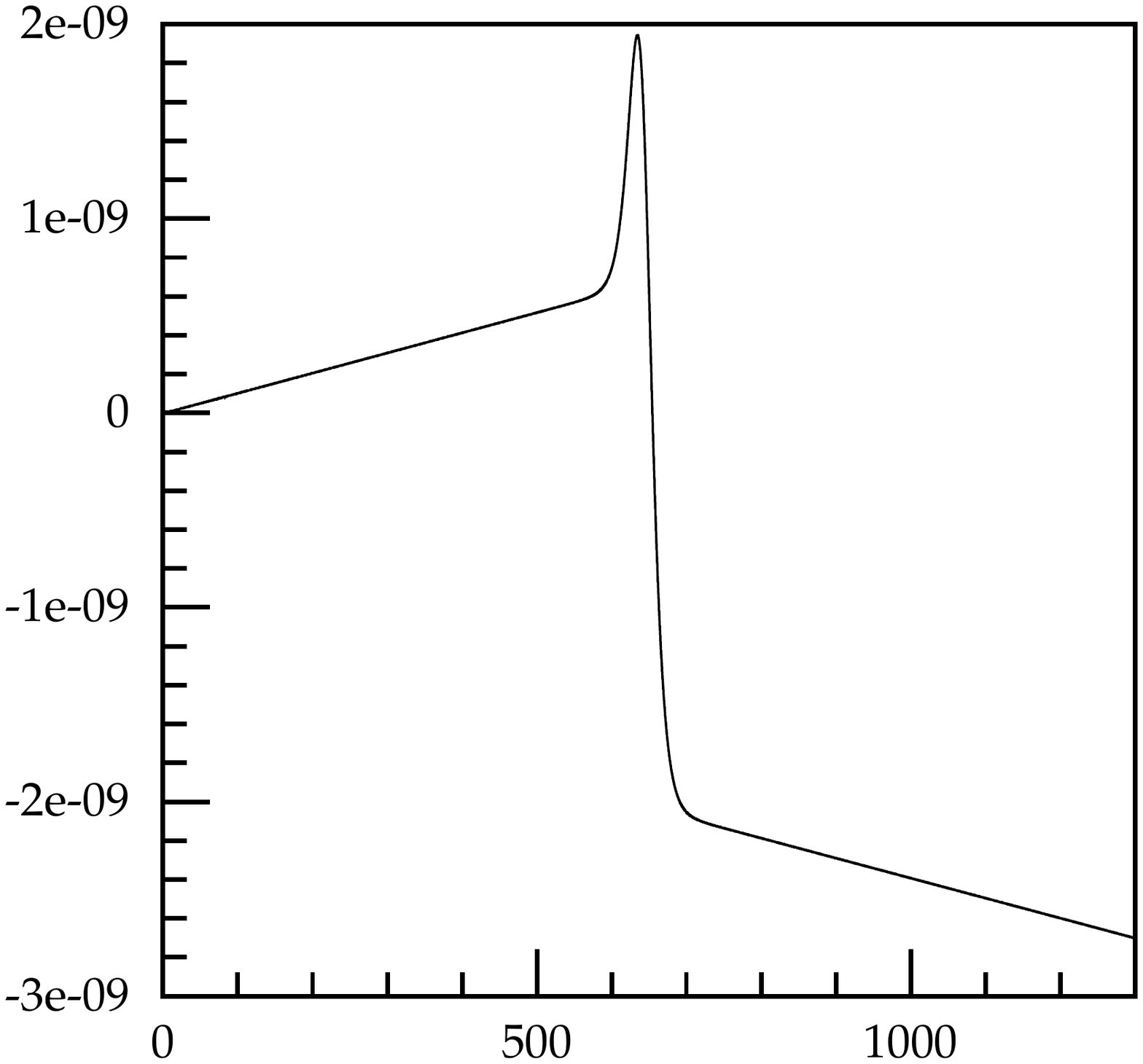}  
\end{center}
\caption[AS]{\small As fig. \ref{fig8} but for $\gamma=0.002$.} 
\label{fig10}
\end{figure}

\begin{figure}
\begin{center}
\includegraphics[width=0.3\textwidth]{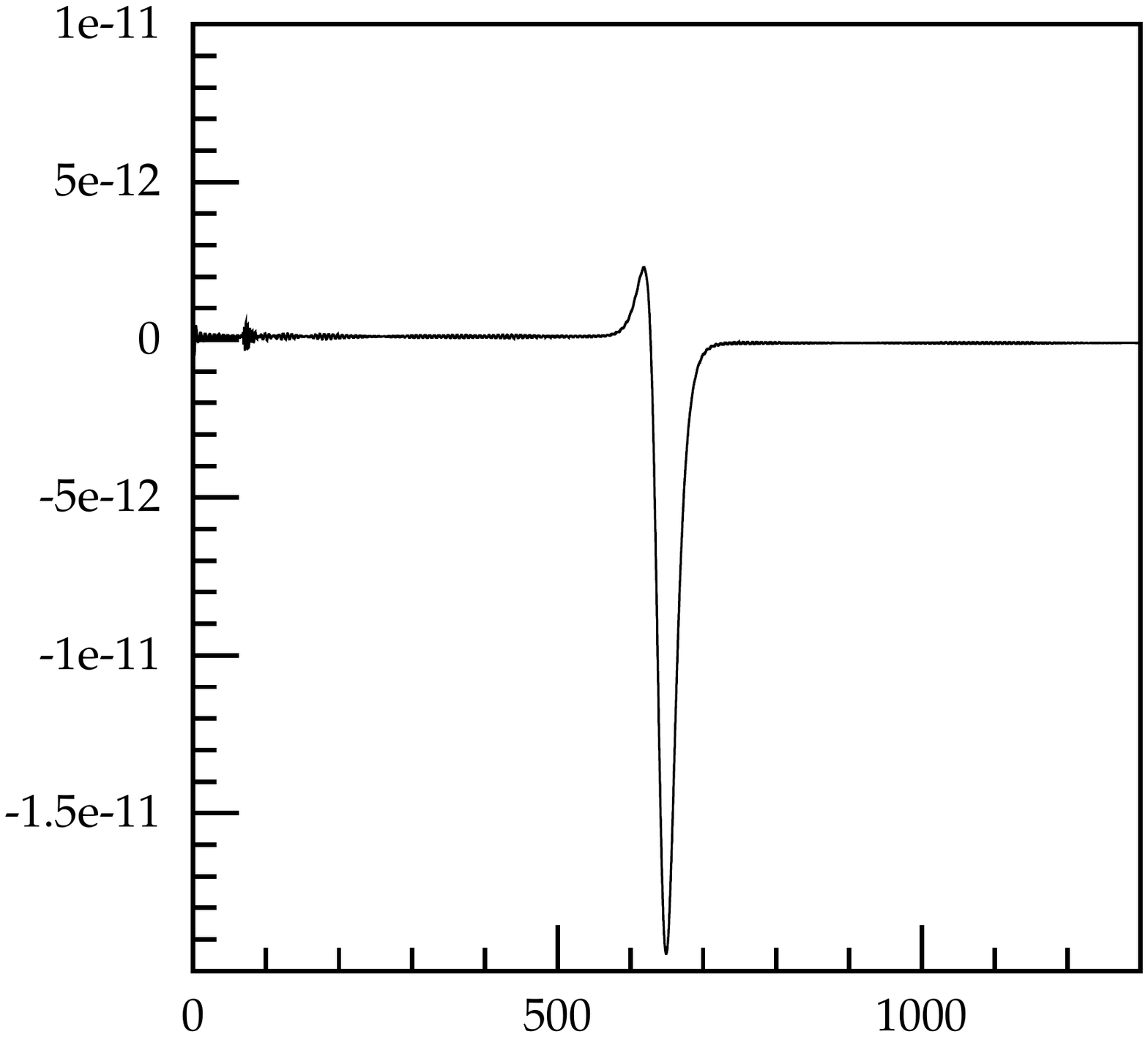} 
\includegraphics[width=0.3 \textwidth]{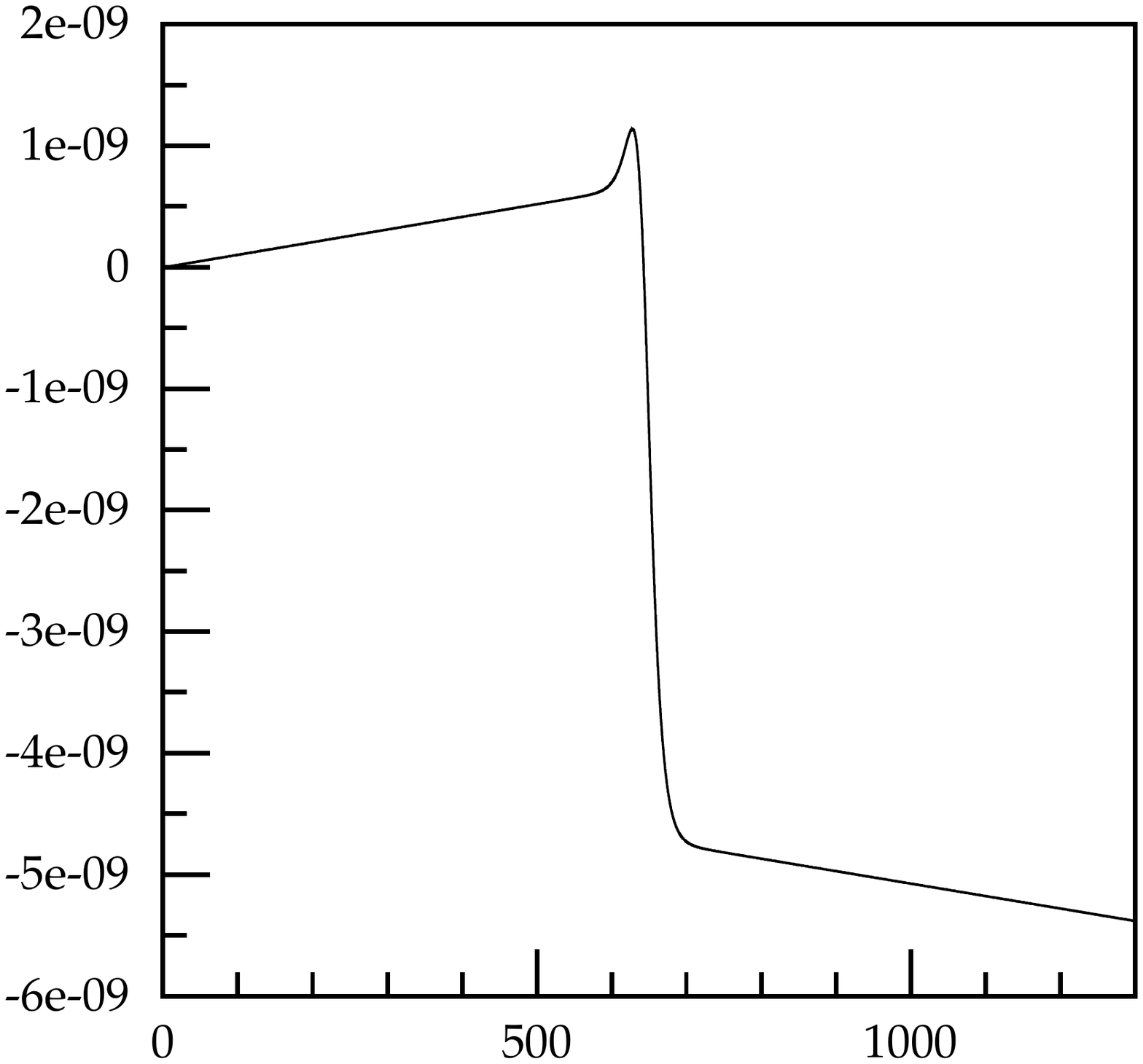}  
\end{center}
 \caption[AS]{\small As fig. \ref{fig8} but for $\gamma=0.004$} 
\label{fig11}
\end{figure}

We then calculated the anomalies in the third class models and performed many simulations for many values of $\varepsilon$ and $\gamma$.
 For most values 
the results were qualitatively the same as in the previous models but for ranges of parameters $\varepsilon$ and $\gamma$ we observed
the transfer of behaviour from the model with approximate symmetry to the one in which this symmetry is broken.
In fig. \ref{fig8}-\ref{fig11} we illustrate this transfer. As we increase the values of $\gamma$ the unintegrated anomaly becomes asymmetric
leading to the change of total anomaly. Figures \ref{fig9} and \ref{fig10} show that this symmetry can be broken in either direction 
(depending on the sign of $\gamma$).

\begin{figure}
\begin{center}
\includegraphics[width=0.3\textwidth]{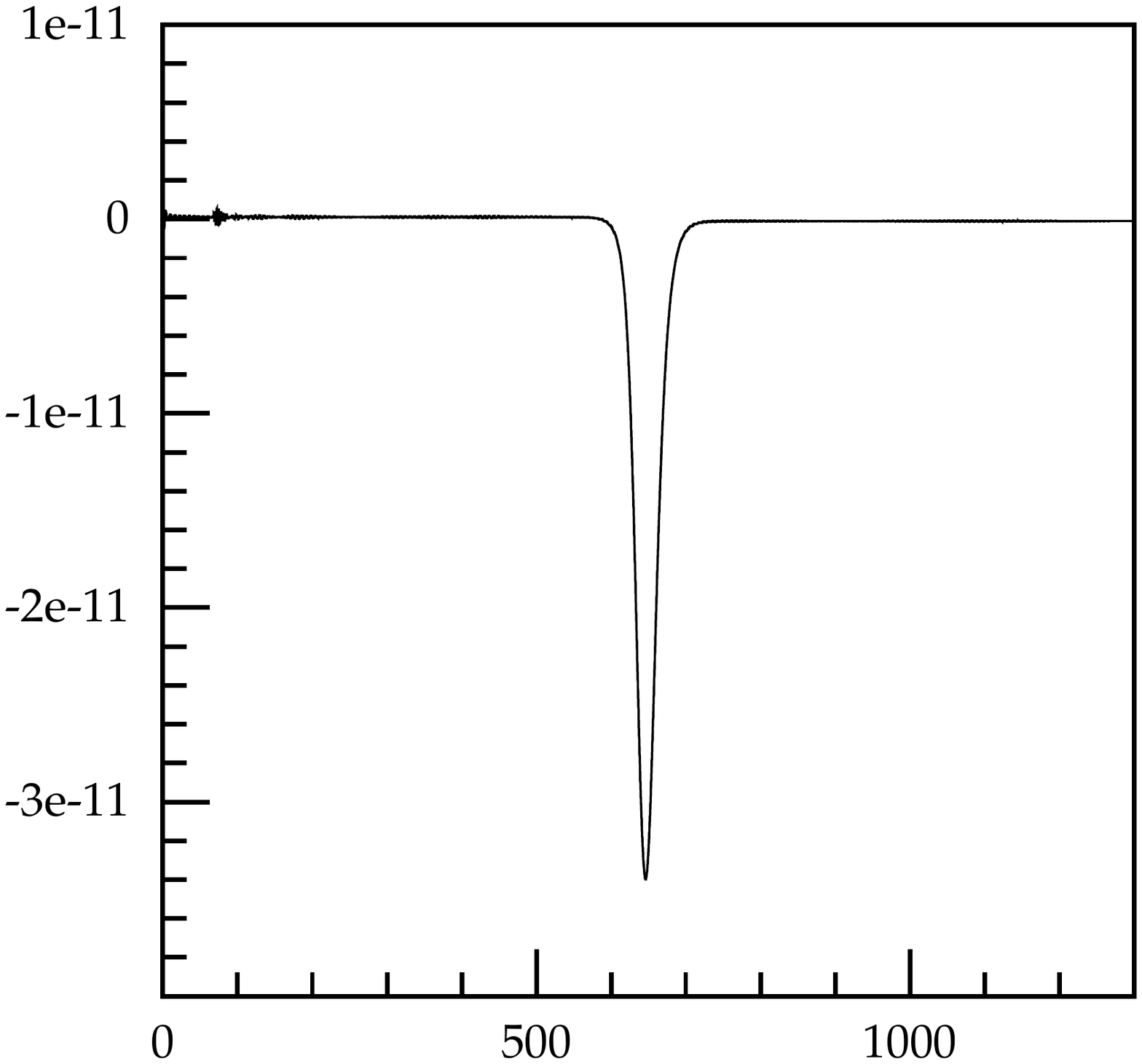} 
\includegraphics[width=0.3 \textwidth]{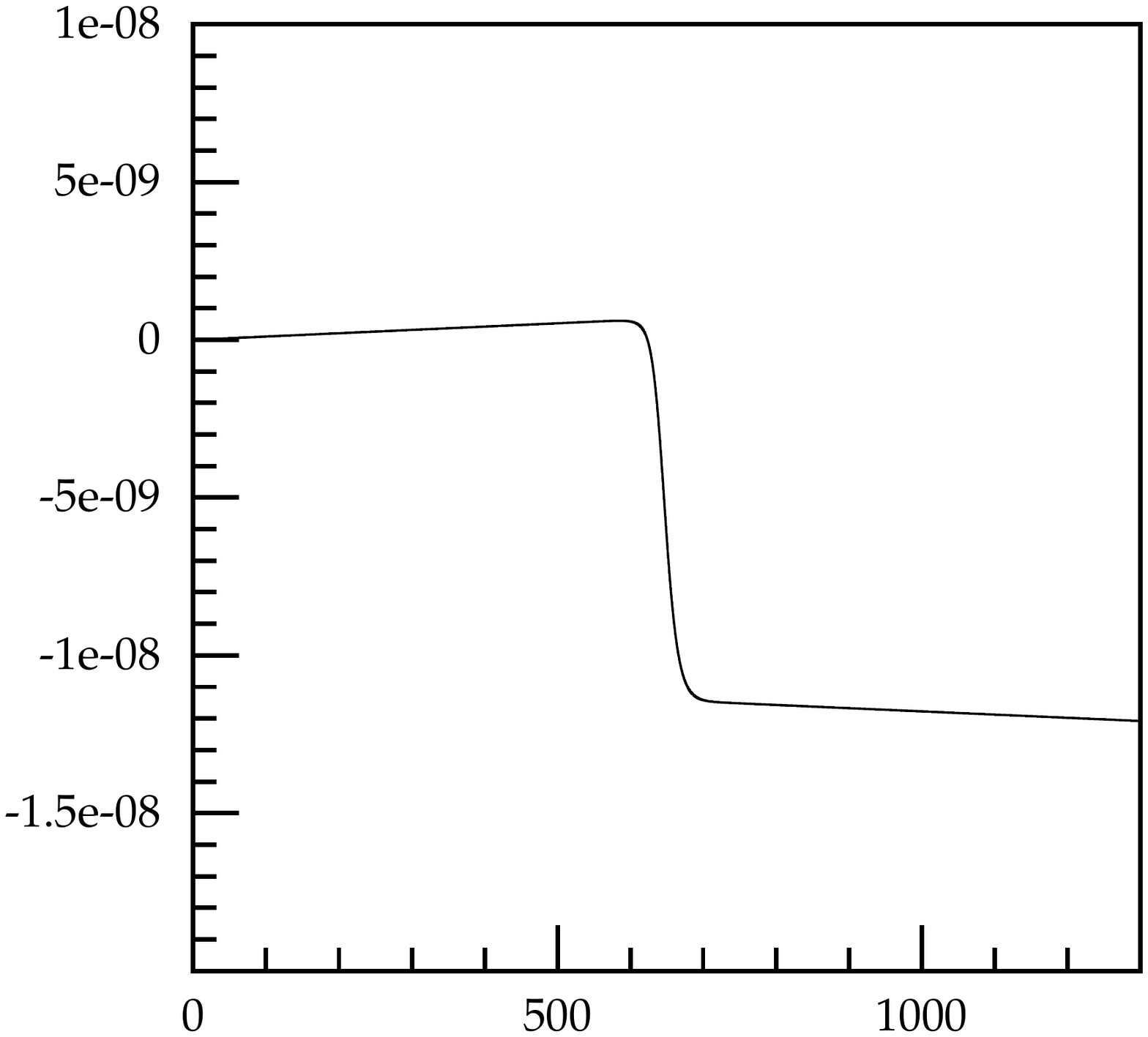}  
\end{center}
 \caption[AS]{\small As fig. \ref{fig8} but for $\gamma=0.009$} 
\label{fig12}
\end{figure}

\subsection{Breather-like configurations}

In all two soliton configurations the interactions between the two solitons were very short-lived; the solitons interacted with each other only when they were very closed to each other and after the reflection their interactions rapidly decreased.
Our numerical results show this very clearly; anomaly changes significantly when the solitons are close together. 

On the other hand when we consider breather-like configurations ({\it i.e.} trapped pairs of a soliton and an antisoliton) they constantly 
interact with each other and so the anomalies change all the time.
Whether their behaviour is consistent with our anomaly behaviour for breather like objects \rf{mirrorcharge} can be studied numerically too
and we have performed several simulations in the third class of models.

To start the simulations we took a breather configuration for the sine-Gordon model \rf{breathersol} and then perfomed the change
of variables \rf{three.one} to obtain the corresponding $\phi$ field. We then used  this field  and its derivative at $t=0$
as the initial conditions for the simulations.

We have performed several simulations for different values of the frequency of the breather {\it i.e.} $\nu$ in \rf{breathersol}
and for various values of $\varepsilon$ and $\gamma$. 

\begin{figure}
\begin{center}
\includegraphics[width=0.3\textwidth]{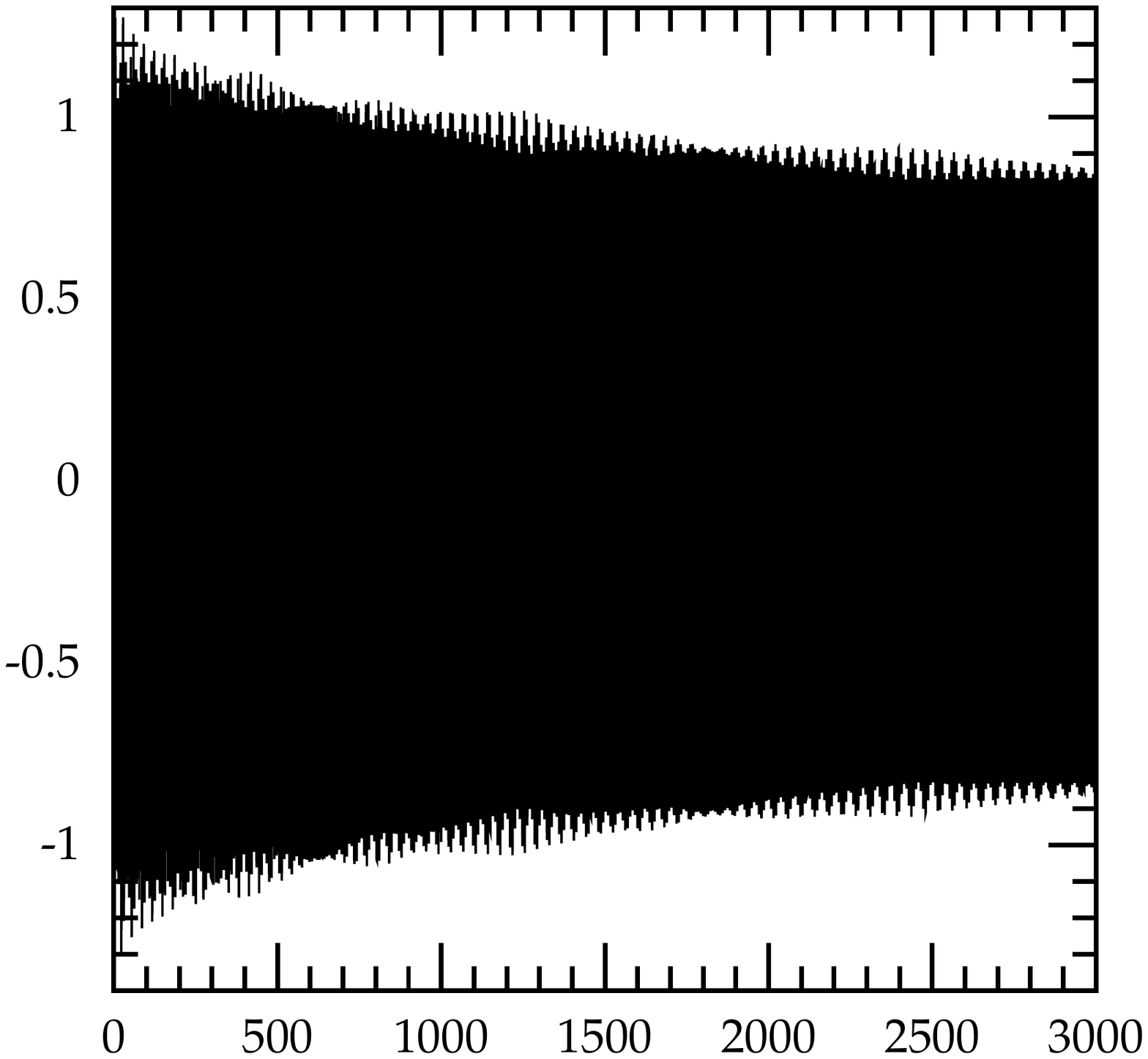} 
\includegraphics[width=0.3 \textwidth]{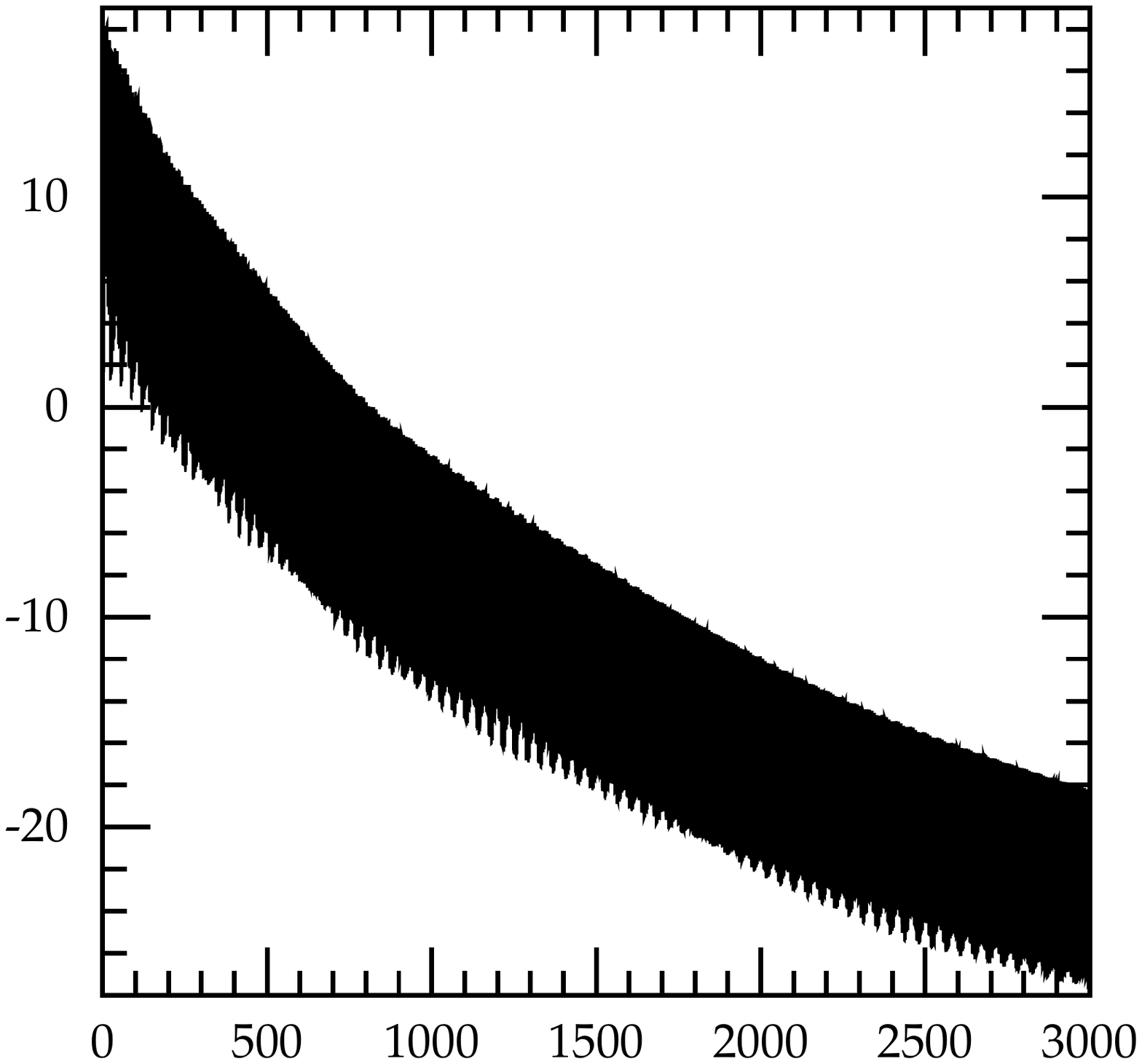}  
\end{center}
\caption[AS]{\small Anomaly (a) and time integrated anomaly (b) for breather-like configurations seen in the third class of models for $\varepsilon=0.1$ and $\gamma=-0.5$.} 
\label{fig13}
\end{figure}

\begin{figure}
\begin{center}
\includegraphics[width=0.3\textwidth]{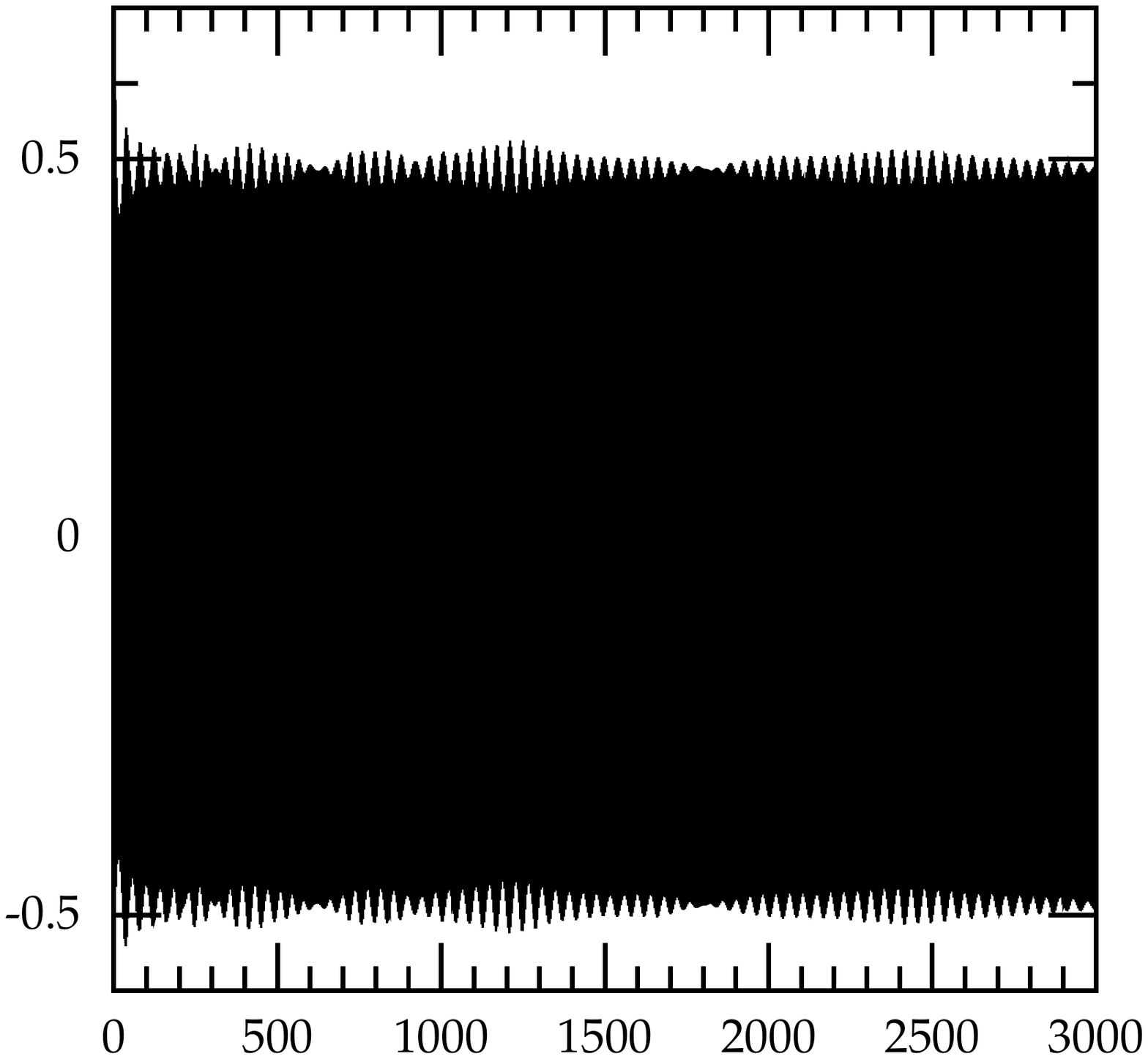} 
\includegraphics[width=0.3 \textwidth]{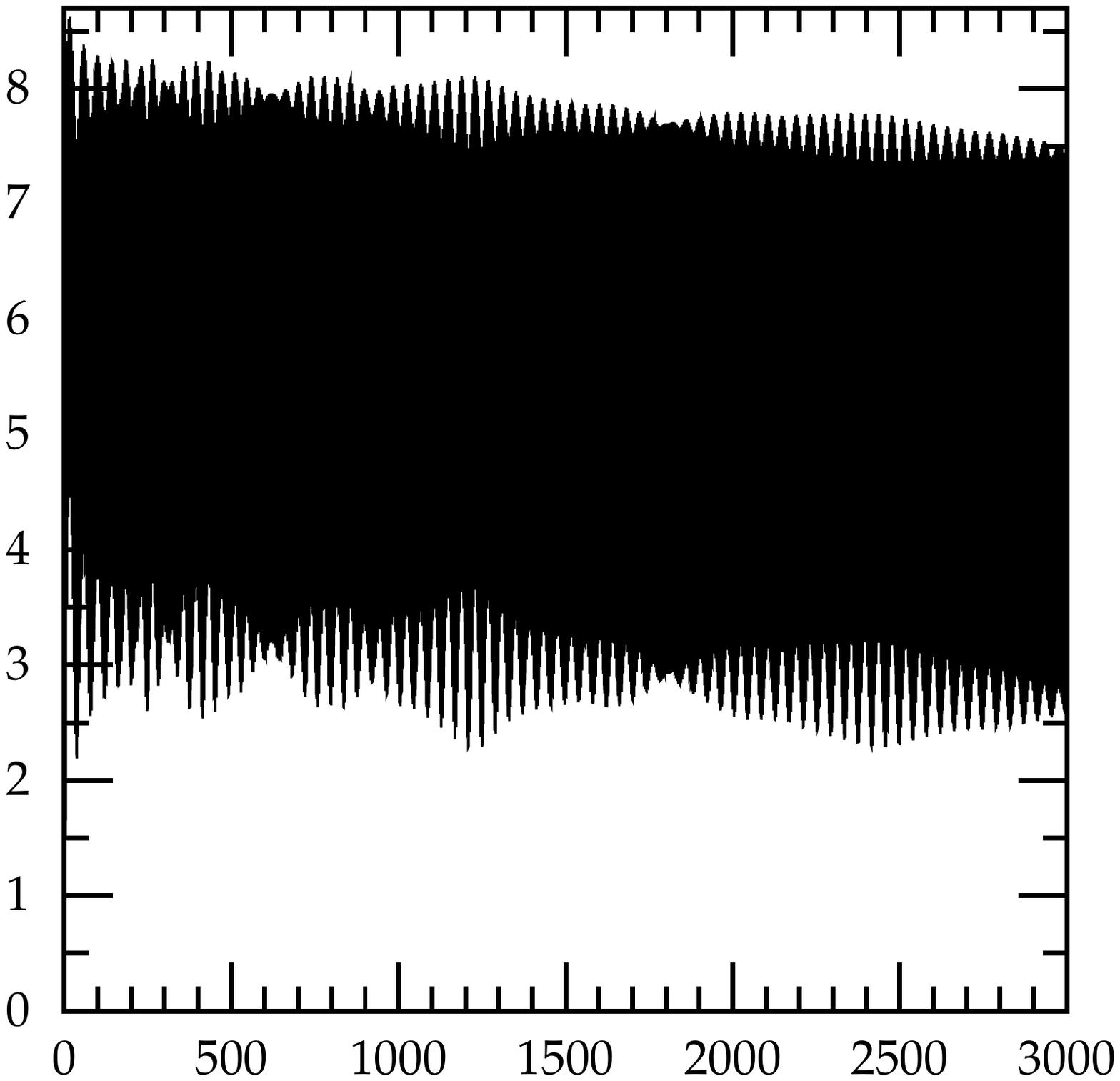}  
\end{center}
\caption[AS]{\small As fig. \ref{fig13} but for $\gamma=0.000$.} 
\label{fig14}
\end{figure}

\begin{figure}
\begin{center}
\includegraphics[width=0.3\textwidth]{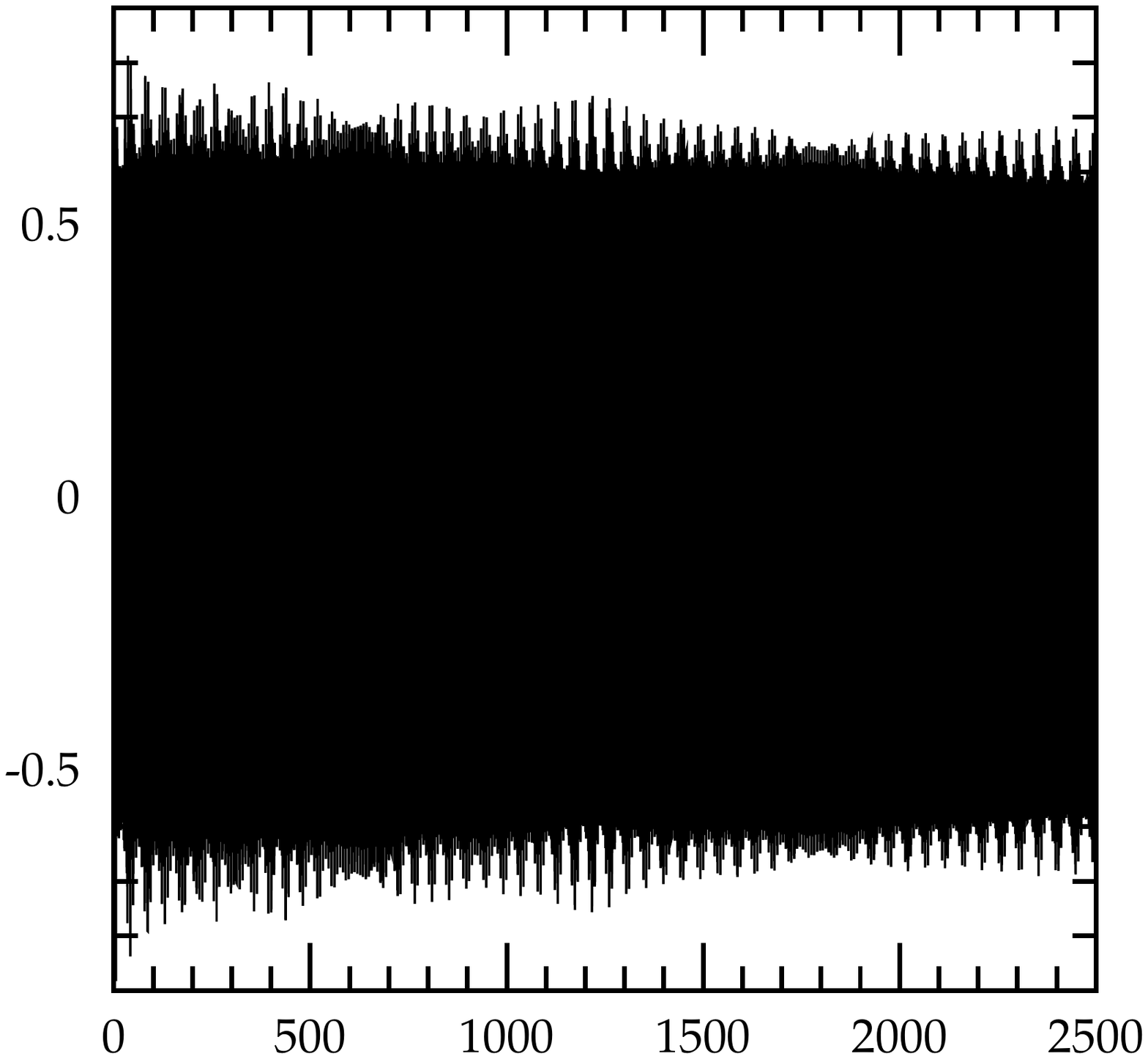} 
\includegraphics[width=0.3 \textwidth]{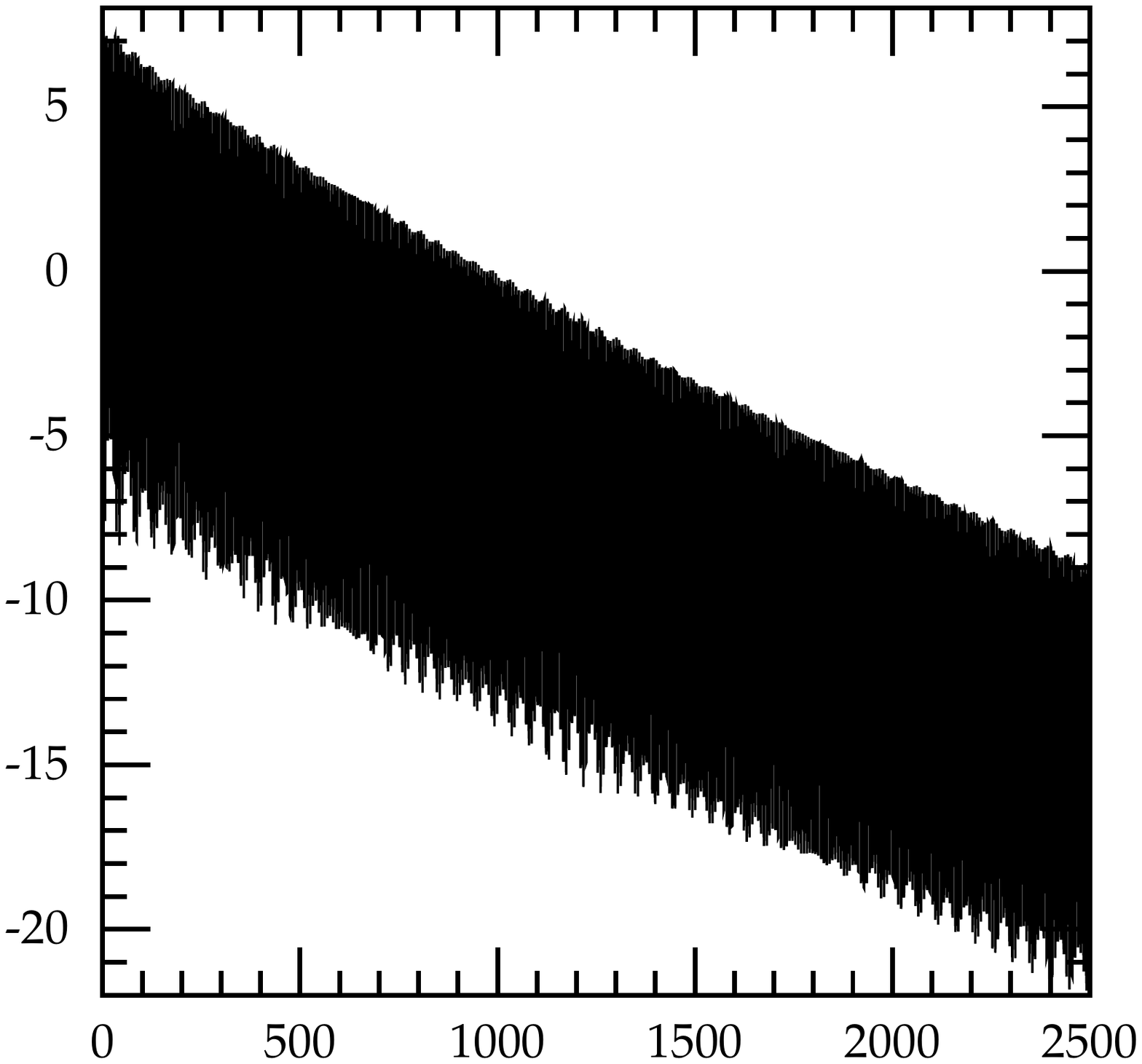}  
\end{center}
 \caption[AS]{\small As fig. \ref{fig13} but for $\gamma=0.5$.} 
\label{fig15}
\end{figure}

Most of our simulations were performed for $\nu=0.5$. These simulations have shown that, indeed, for $\gamma=0.00$ we have quasi-integrability, {\it i.e.} the behaviour is consistent with \rf{mirrorcharge}. The anomaly does change all the time and even the time 
integrated anomaly does depend on time. However, the oscillations are approximately around a fixed value and so, shifting in time and choosing $t_{\Delta}$ appropriately we see that \rf{mirrorcharge} holds.
When we take small $\gamma$ the effects are not very different ({\it i.e.} one should wait long enough to see the difference 
but for larger values of $\gamma$ the effects are clear
 as the time integrated anomaly decreases.
This is true for positive and negative values of $\gamma\ne0$. 
In fig \ref{fig13}, \ref{fig14} and \ref{fig15} we present results of our simulations for $\nu=0.5$ for $\epsilon=0.1$ and three values of $\gamma$, namely 
$\gamma=-0.5$ (fig \ref{fig13}), $\gamma=0.0$ (fig \ref{fig14}) and $\gamma=0.5$ in fig \ref{fig15}.
The left hand plots in each set of pictures show unintegrated anomaly, the right hand sides ones the time integrated ones.

\begin{figure}
\begin{center}
\includegraphics[width=0.3\textwidth]{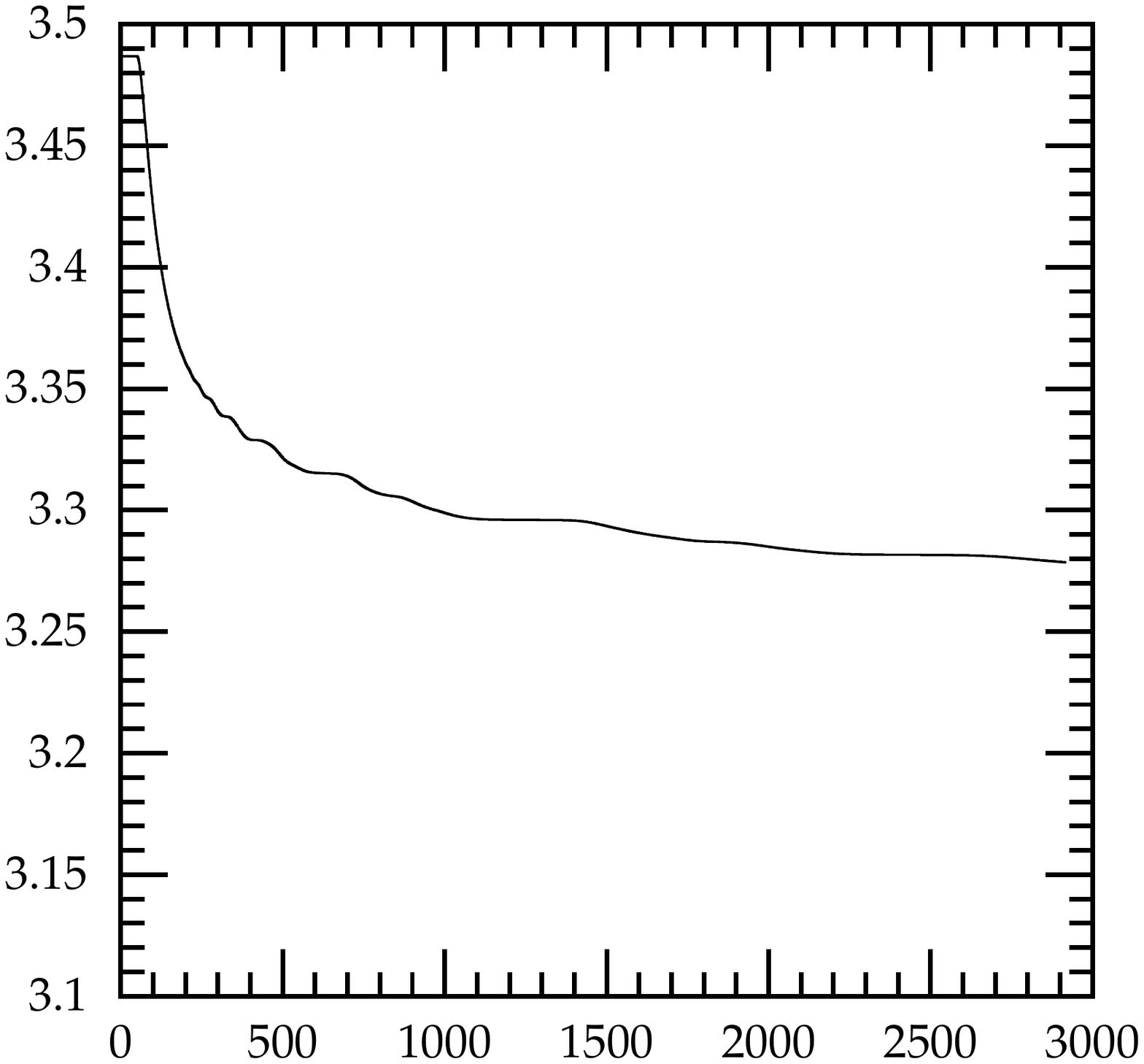} 
\includegraphics[width=0.3 \textwidth]{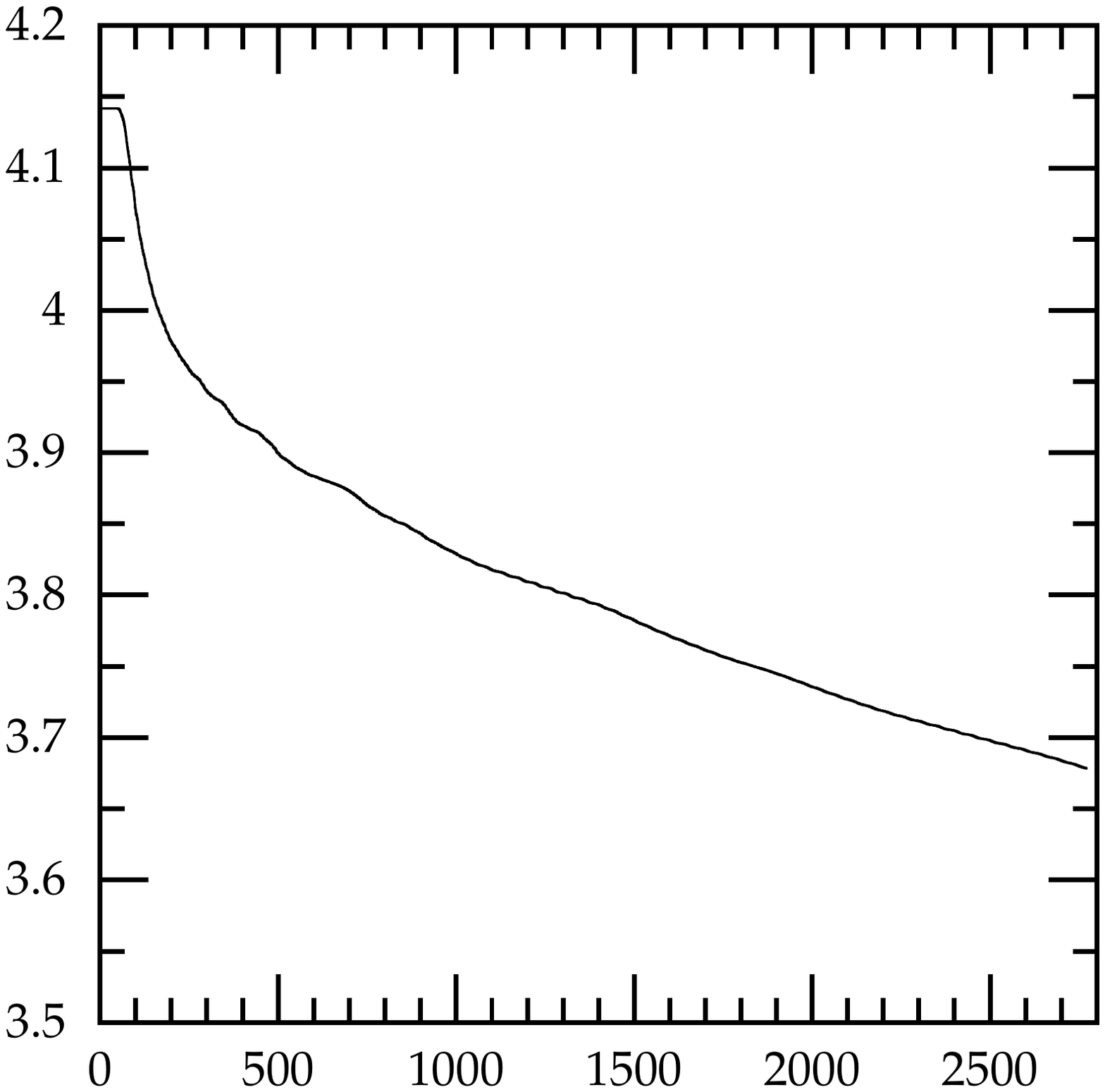}  
\end{center}
 \caption[AS]{\small Total energy seen for $varepsilon=0.1$ and $\gamma=0.0$ (left) and $\gamma=0.5$ (right plot) } 
\label{fig16}
\end{figure} 

The presented results clearly show that the $\gamma=0$ case is somewhat special as then the time integrated anomaly exibits the expected 
symmetry principle \rf{mirrorcharge}.

Of course, as is well known the modified sine-Gordon models do not possess breather solutions except for the sine-Gordon case.
Thus, for all values of $\varepsilon$ the breather-like structures we have looked at gradually decay. This decay is carried out by the emission of radiation during the repetitive scattering of soliton and antisoliton forming the breather ({\it i.e.} the breathing of the 
breather-like configuration). Clearly, the speed of this decay, depends on $\varepsilon$ and $\gamma$. It also depends on $\nu$ - the frequency of the breather-like structure. We have studied this dependence and have observed that for (for a given $\nu$) the energy of the configuration decreases with the increase of $\varepsilon$ (it decreases by a factor of 3 from $\varepsilon=0.1$ to $\varepsilon=0.5$).
At the same time, for a given $\varepsilon$ and $\nu$ it increases with the increase of $\gamma$ and, like for the pure sine-Gordon model,
for a given $\varepsilon$ and $\gamma$ it decreases with the increase of $\nu$.

These dependences have very important implications for the decay of the studied breather-like structures. Thus, in particular, if one keeps
$\nu$ and $\varepsilon$ fixed the increase of $\gamma$ speeds up the decay of the structure, which for low values of $\gamma$ takes place gradually, while above a certain value of $\gamma$ happens quite suddenly. The value at which this happens depends on $\varepsilon$ and decreases with the increase of $\gamma$.

In fig. 19 we present some preliminary plots of the decrease of the energy of two such breather-like configurations 
(both for $\nu=0.5$ and $\varepsilon=0.1$;  one for $\gamma=0.0$ (the left hand-one) and one for $\gamma=0.5$.

They show the faster decrease of the configuration with $\gamma\ne0$.

\section{Further Comments and Some Conclusions}
\label{sec:conclusions}
\setcounter{equation}{0}

In this paper we have performed further studies of quasi-integrability - based on the observation \cite{us}, \cite{us2} that the anomaly, which distinguishes integrable models
from nonintegrable ones, also vanishes in some nonintegrable models in which the field configurations possess the parity  symmetries discussed in Section 3.

This observation was originally made in some very specific models and here we have tried to assess its general validitity.  
So, we  constructed three  classes of models (one with symmetry, one without it and one (dependent on two parameters) which would allow us to interpolate between the two. Our results have confirmed the validity of our assumption 
(and so extended the class of models in which our observation holds) and have also allowed us to study the way the anomaly varies as we move away from the models with this extra symmetry.

Our work has also led us to look more clearly at breather-like configurations. Such configurations depend on many parameters
and in this paper we have concentrated our attention at looking at the behaviour of such configurations when the symmetry is present and when it is absent. However, the symmetry is only one interesting topic to investigate for the breather-like
configurations. One could look at the aspects of the decay of such configurations and the dependence of this behaviour
of various parameters of these configurations. We plan to look at this next and we hope to be able to report on this in not too distant future.

\vspace{2cm}

{\bf Acknowledgements:} LAF is grateful for the hospitality at the Department of Mathematical Sciences of Durham University where part of this work was carried out.
LAF and WJZ would like to thank the Royal Society for awarding them a grant which made their collaboration possible.  LAF is partially supported by CNPq (Brazil). 
This paper was finished when the authors
were ``researching in pairs'' at the Matematisches Forschunginstitute in Oberwolfach. They would like to thank the Oberwolfach Institute for its hospitality.\\

\vspace{1cm}

\appendix

\section{The algebra}
\label{sec:appendix-algebra}
\setcounter{equation}{0}

We consider the $sl(2)$ algebra
\be
\sbr{T_3}{T_{\pm}}=\pm\, T_{\pm}, \qquad\qquad \sbr{T_{+}}{T_{-}}=2\, T_3.
\ee
We take the following basis for the corresponding loop algebra
\be
b_{2m+1}=\lambda^{m}\(T_{+}+\lambda\, T_{-}\),\qquad 
F_{2m+1}=\lambda^{m}\(T_{+}-\lambda\, T_{-}\),\qquad
F_{2m}=2\,\lambda^m\, T_3.
\ee
The algebra is
\br
\sbr{b_{2m+1}}{b_{2n+1}}&=&0,\nonumber\\
\sbr{F_{2m+1}}{F_{2n+1}}&=&0,\nonumber\\
\sbr{F_{2m}}{F_{2n}}&=&0,\nonumber\\
\sbr{b_{2m+1}}{F_{2n+1}}&=&-2\, F_{2(m+n+1)},\nonumber\\
\sbr{b_{2m+1}}{F_{2n}}&=&-2\, F_{2(m+n)+1},\nonumber\\
\sbr{F_{2m+1}}{F_{2n}}&=&-2\, b_{2(m+n)+1}.\nonumber
\er

We have a grading operator
\be
d= T_3+ 2\,\lambda \frac{d\;}{d\lambda}
\lab{gradingop}
\ee
such that
\be
\sbr{d}{b_{2m+1}}=\(2m+1\)\,b_{2m+1},\qquad\qquad 
\sbr{d}{F_{m}}=m\,F_{m}.
\ee

\end{document}